\documentclass[sigconf,natbib,screen=true,review=false]{acmart}

\usepackage[skip=0pt]{caption}
\usepackage{booktabs}
\usepackage{numprint}
\npthousandsep{,}
\npdecimalsign{.}
\usepackage[inline]{enumitem}
\usepackage{balance}

\usepackage{xcolor}

\usepackage{verbatim}

 \usepackage{multirow}

\usepackage{graphicx} 
\graphicspath{{img/}} 

\usepackage{acronym}
\acrodef{IR}{information retrieval}

\usepackage{subfigure}

\copyrightyear{2020}
\acmYear{2020}
\setcopyright{rightsretained}
\acmConference[SIGIR eCom '20]{The 2020 SIGIR Workshop On eCommerce}{July 30, 2020}{Virtual Event, China}
\acmBooktitle{Proceedings of the 2020 SIGIR Workshop on eCommerce (SIGIR eCom'20), July 30, 2020, Virtual Event, China}
\acmPrice{}

\newcommand{\header}[1]{\vspace{1mm}\noindent\textbf{#1}.}

\AtBeginDocument{%
  \providecommand\BibTeX{{%
    \normalfont B\kern-0.5em{\scshape i\kern-0.25em b}\kern-0.8em\TeX}}}

\newcommand{\justabit}{$\, \, \, \, \,$}

\author{%
Mariya Hendriksen$^1$ \justabit
Ernst Kuiper$^2$ \justabit
Pim Nauts$^3$ \justabit
Sebastian~Schelter$^{4,5}$ \justabit
Maarten~de~Rijke$^{4,5}$
}
\affiliation{%
  \institution{
  $^1$AIRLab, University of Amsterdam \justabit
  $^2$Bol.com\justabit
  $^3$Albert Heijn\justabit
  $^4$University of Amsterdam\justabit
  $^5$Ahold Delhaize 
  }
}  
\email{m.hendriksen@uva.nl, ekuiper@bol.com, pim.nauts@ah.nl, s.schelter@uva.nl, m.derijke@uva.nl}

\def\authors{Mariya Hendriksen, 
Ernst Kuiper, 
Pim Nauts, 
Sebastian Schelter and
Maarten de Rijke}

\setcopyright{rightsretained}
\acmConference[SIGIR eCom'20]{ACM SIGIR Workshop on eCommerce}{July 30, 2020}{Virtual Event, China}
\acmYear{2020}
\copyrightyear{2020}
\makeatletter
\renewcommand\@formatdoi[1]{\ignorespaces}
\makeatother
\acmISBN{}

\begin{document}

\title[Analyzing and Predicting Purchase Intent in E-commerce]{Analyzing and Predicting Purchase Intent in E-commerce:\\ Anonymous vs.\ Identified Customers}

\renewcommand{\shortauthors}{Hendriksen et al.}

\begin{abstract}
The popularity of e-commerce platforms continues to grow. 
Being able to understand, model, and predict customer behaviour is essential for customizing the user experience through personalized result presentations, recommendations, and special offers.
Previous work has considered a broad range of prediction models as well as features inferred from clickstream data to record session characteristics, and features inferred from user data to record customer characteristics.
So far, most previous work in the area of purchase prediction has focused on known customers, largely ignoring anonymous sessions, i.e., sessions initiated by a non-logged-in or unrecognized customer.
However, in the de-identified data from a large European e-commerce platform available to us, more than 50\% of the sessions start as anonymous sessions.

In this paper, we focus on purchase prediction for both anonymous and identified sessions on an e-commerce platform.
We start with a descriptive analysis of purchase vs. \ non-purchase sessions. 
This analysis informs the definition of a feature-based model for purchase prediction for anonymous sessions and identified sessions; our models consider a range of session-based features for anonymous sessions, such as the channel type, the number of visited pages, and the device type. 
For identified user sessions, our analysis points to customer history data as a valuable discriminator between purchase and non-purchase sessions.
Based on our analysis, we build two types of predictors:
\begin{enumerate*}
\item a predictor for anonymous sessions that can accurately predict purchase intent in anonymous sessions, beating a production-ready predictor by over 17.54\% $F_{1}$; and 
\item a predictor for identified customers that uses session data as well as customer history and achieves an $F_{1}$ of 96.20\% on held-out data collected from a real-world retail platform.
\end{enumerate*}
Finally, we discuss the broader practical implications of our findings.
\end{abstract}




\maketitle


\section{Introduction}
\label{sec:introduction}

Information retrieval (\acs{IR}\acused{IR}) technology is at the heart of today's e-commerce platforms, in the form of search engines, recommenders, and conversational assistants that connect users to the products they may be interested in~\citep{rowley2000product}.
To help improve the effectiveness of \ac{IR} technology in an e-commerce context, the problem of analyzing, modeling, and, ultimately, predicting customers' purchase intent has been studied extensively in academia and industry~\citep{bellman1999predictors, agichtein2006learning, lo2016understanding}

\header{Purchase intent prediction} Here, purchase intent is defined as a predictive measure of subsequent purchasing behavior~\citep{morwitz1992using}.

\begin{figure}[h]
\centering
      \includegraphics[clip,trim=0mm -5mm 0mm 0mm,width=0.8\columnwidth]{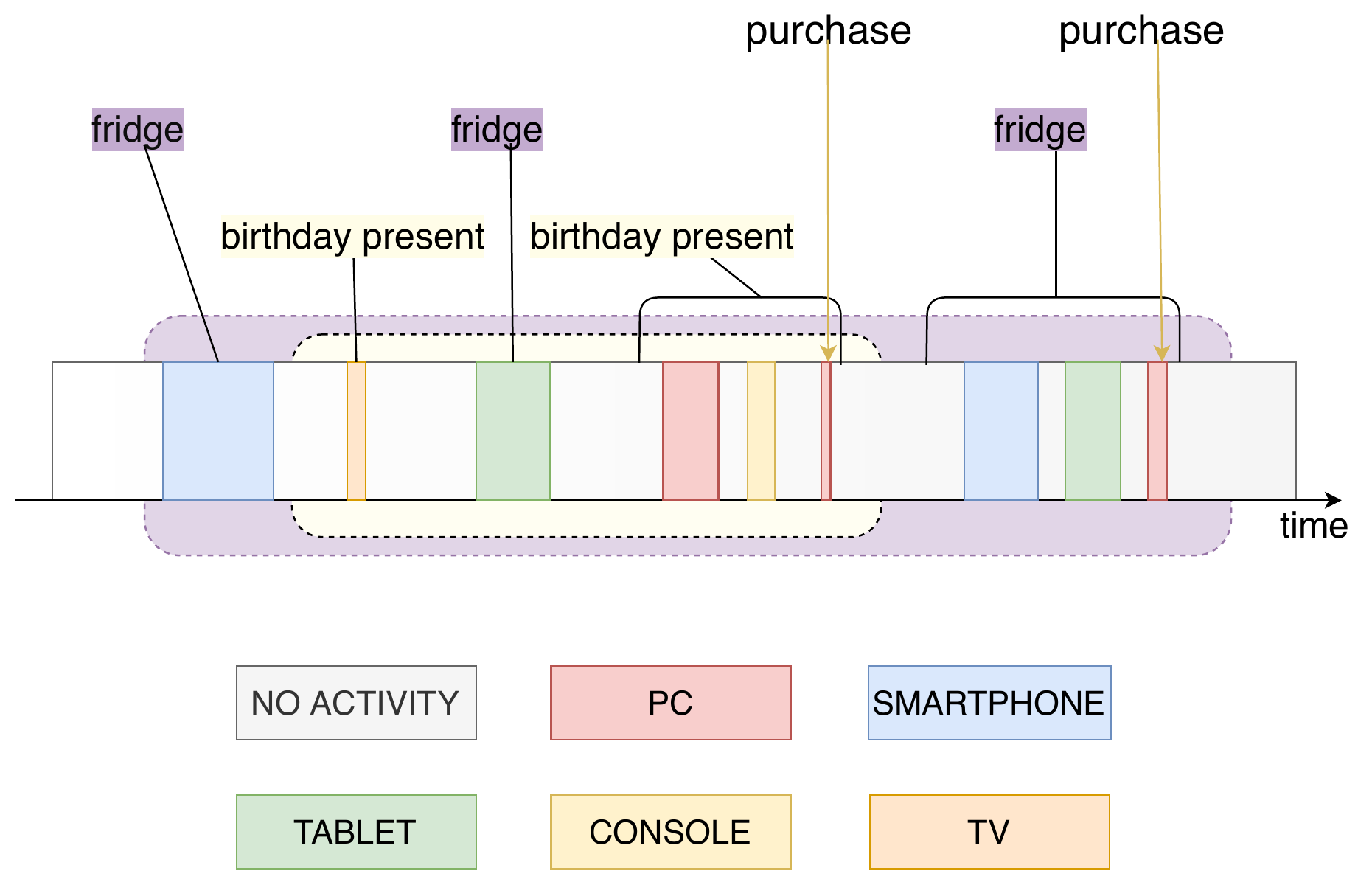}
  \caption{Customer journeys across sessions, with multiple interests and devices; the colors indicate different devices.}
  \label{fig:teaser}
\end{figure}

Figure~\ref{fig:teaser} illustrates the complexities of customer behavior during a sequence of sessions, when multiple tasks, interests, and devices may play a role. 
Areas in the back of the figure are meant to signify different user journeys across time, purple for one that is focused on fridges, yellow for one that is focused on a birthday present. Colored rectangular blocks in the front indicate different devices used by the user. Initial exploration of a relatively expensive item (a fridge) starts on a smartphone and continues on a tablet, while the journey ends with a purchase of a fridge on a PC.
The purchase of a fridge is interleaved with the purchase of a (lower-priced) birthday present, with initial exploration on a PC, followed by further exploration on a TV and PC, and, ultimately, a purchase on a PC.

Online search behavior that targets transactions has been analyzed at scale at least since the work by~\citet{broder-2002-taxonomy}, who identified a class of so-called \emph{transactional} queries, where the user is seeking to reach a page with more interaction opportunities, e.g., to conduct a purchase, download or sign-up. In particular, factors influencing online purchases have been described as early as in 2002~\citep{george2002influences}, and work on predicting purchases goes back to at least to the work of~\citep{ben-shimon-2015-recsys}, where the task was to predict whether a given customer is going to purchase within a given session.

\header{Challenges} Despite the many advances, purchase intent prediction still has many challenges~\citep{tsagkias-2020-challenges}. In particular, previous work on purchase intent prediction has focused mostly on customers of an e-commerce platform who are identified or recognized by the platform. A diverse range of models has been considered, from traditional feature-based models such as boosted decision trees to sequence-based neural models such as RNNs.
However, based on the analysis of de-identified data from an e-commerce website available to us, more than 50\% of traffic comes from anonymous users. Purchase intent detection for anonymous users is particularly challenging because it cannot rely on historical information about the user on which many of the existing models rely.

\header{Features for purchase intent prediction} In this paper, we focus on identifying signals that suggest purchase intent in an anonymous and identified setting. We do this by analyzing purchase vs. non-purchase sessions sampled from a large European e-commerce website and testing the features based on our observations on a production-ready model. We further test the obtained feature sets on five other classifiers to explore the generalizability of our findings.
In particular, we include features derived from session-based data such as page dwell time and customer-specific data such as the number of days since the last purchase. Session-based features have the advantage that they are available both during sessions when a user is identified (i.e., the customer has logged-in or is recognized through cookies) and anonymous sessions (when the customer is not known). Customer-related features are only available during identified sessions.
Interestingly, many of the features proposed previously~\citep{seippel-2018-customer} apply only to identified sessions: \emph{purchase intent prediction for anonymous sessions has been studied very little}.

To fill this gap, we analyze a dataset of more than 95 million sessions, sampled from four weeks of anonymized user interaction data in a European e-commerce platform. We answer the following research questions:

{\em RQ1: How do purchase sessions differ from non-purchase sessions?} In Section~\ref{section:characterizingpurchaseintent} we compare purchase vs. \ non-purchase sessions in such aspects as session length, temporal variations, device and channel type, queries. Among others, we find out that purchase sessions tend to be longer than non-purchase ones, customers are more likely to purchase in the evening and during a weekday, and more likely to own more than 1 device.

{\em RQ2: What are the important session-based features that allow us to tell purchase sessions apart from non-purchase sessions? What are the important historical features that should inform predictors for identified sessions? How does the importance of features change across the session?} Based on the experiments described in Section~\ref{section:predictingpurchaseintent}, we conclude that historical features related to previous purchasing behavior are highly important for detecting purchases in the identified setting.  For the anonymous setting, however, dynamic features related to page dwell time and sequence of pages are most important. Besides, the importance of dynamic features increases as the session continues, while the importance of static features decreases.

{\em RQ3: How effective are models used for purchase intent prediction for anonymous vs.\ identified sessions? Furthermore, to which degree do the proposed features help improve performance for anonymous sessions?} In Section~\ref{section:predictingpurchaseintent}, we show that in the anonymous setting, tree-based and neural classifiers demonstrate the best performance, and adding extra features to models improves $F_{1}$ by about 17\%. In contrast, for identified setting all models demonstrate high performance and adding extra features do not provide a significant gain.

\noindent%
 
The principal contributions of our research are the following:
\begin{itemize}[leftmargin=*,nosep]
   \item We conduct an in-depth analysis of a real-world customer interaction dataset with more than 95 million sessions, sampled from a large European e-commerce platform. We identify session features such as device type and conversion rate, weekday, channel type, and features based on historic customer data such as number of previous orders and number of devices to distinguish between purchase and non-purchase sessions (see Section~\ref{section:characterizingpurchaseintent}).
   \item We define two feature sets for purchase prediction, tailored towards anonymous sessions and identified sessions (see~Section~\ref{section:predictingpurchaseintent}).
   \item We evaluate our proposed features by extending an existing production-ready model and run additional experiments with classifiers generally used for this task. We find  $F_{1}$ improvements of up to 17\% in purchase intent prediction for anonymous sessions and reach an $F_{1}$ of 96\% for identified sessions on held-out data collected from a real-world retail platform (see ~Section~\ref{section:predictingpurchaseintent}).
\end{itemize}


\section{Background and Definitions}
\label{sec:backgroundanddefinitions}

In our study, we operate with the following definitions.

A \textit{session} is a sequence of requests made by a single end-user during a visit to a particular site. A session ends if the user is idle for more than 30 minutes.
We define two types of sessions: \textit{purchase sessions}, during which the customer buys an item, and \textit{non-purchase sessions}, during which the customer does not buy anything.
In connection to this, we define \emph{purchasers} as customers who had at least one purchase session, whereas \emph{non-purchasers} are customers who were identified but have never purchased anything.
We furthermore distinguish between \textit{identified sessions}, where a customer is logged in or recognized with a browser cookie, and \textit{anonymous} sessions where this is not the case. 
Additionally, we denote the number of actions taken during a given session as the \textit{session length}, where an \textit{action} corresponds to opening a new web page, submitting a search query, or adding/removing an item to/from the shopping basket.

\textit{Device switch} is the act of changing the type of browsing device between two consecutive sessions that belong to the same journey. 
For instance, if a customer first explores the platform on a smartphone and afterward accesses the platform on a PC, she switches from a smartphone to a PC.

A \textit{channel} indicates the way through which a customer enters the platform. 
For example, if the customer comes to the platform via an advertisement, she uses a paid channel.

The \textit{conversion rate} denotes the fraction of visits during which a purchase was made~\citep{moe2004dynamic}. 
We use this metric to compare device popularity in a purchasing context. 
We calculate the conversion rate by dividing the number of purchasing sessions by the overall number of sessions. 
In order to protect sensitive information, we only report \textit{standardized conversion rates} for each device; since we are interested in differences across devices types, this suffices for our purposes. 
The standardized conversion rate is computed by subtracting the mean conversion rate per device type from the desired conversion rate and dividing the result by the standard deviation of the device-specific conversion rate. 
For instance, if our device specific conversion rates are
$\text{\em Conversion Rates} = \{0.5, 0.2, 0.3\}$, the mean of device-specific conversion rates is $\overline{\text{\em Conversion Rate}} = 0.33$ and the standard deviation of conversion rates is $\sigma = 0.12$. 
Therefore, the resulting standardized conversion rates are $\text{\em Standardized Conversion}$ $\text{\em Rates} = \{1.34, -1.07, -0.27\}$

\section{Dataset Description}
\label{sec:dataset}

In this section, we describe how we extract a dataset consisting of anonymized user interaction data from the search logs of an e-commerce platform, and summarize dataset statistics.

\header{Data Collection} Our dataset comprises four weeks (28 days) of anonymized visits sampled from a European e-commerce platform in October 2019. 
The original sample of the log entries includes a unique non-personal customer identifier (for identified users), the type of browsing device used during the session, as well as a timestamp for every query, and a URL of each clicked page. 
We convert all the timestamps to the Central European Time Zone (CET). 
We additionally recorded the price of every product the customers have seen and the prices of the items they ended up buying. 
In cases where a customer starts a session without logging in and ends up logging in at a later point in the session, we assign the session to the customer.

To filter out bot traffic, we apply several measures related to location and device type~\citep{bomhardt-2005-web}. 
First, we filter out sessions based on location, to only include entries from the European countries from which the majority of the customers come; bots come mostly from non-European IPs, especially North-America.  
Second, we specify the set of device types we are interested in and remove all the entries from other devices, leaving us with \emph{PC}, \emph{Smartphone}, \emph{Tablet}, \emph{Game Console}, and \emph{TV}; bots often do not specify a device type.

\header{Dataset Statistics} Table~\ref{tab:ds_stats} provides descriptive statistics of the resulting dataset. 
Overall, the dataset contains \numprint{95757177} sessions, out of which \numprint{54144152} (about 56.5\%) are anonymous. 
In total, the dataset contains \numprint{9663509} identified users. 
We additionally keep track of the device types used for browsing and distinguish between five such device types: PC, smartphone, tablet, game console, and TV.
The table also lists the number of search queries; these are the queries submitted during the sessions captured in the log. 

\begin{table}[!htb]
  \caption{Dataset statistics.}
  \label{tab:ds_stats}
  \begin{tabular}{lr}
    \toprule
    \bf Description &  \bf Total \\
    \midrule
    Sessions        &  \numprint{94402590} \\
    \quad Anonymous                  &  \numprint{55305709} \\
    \quad Logged-in or recognized                  &  \numprint{39096881} \\
    Logged-in or recognized customers         & \numprint{6125781} \\
    Queries    & \numprint{31185176} \\
    \multirow{2}{*}{Device types} & \multicolumn{1}{l}{PC, Smartphone, Tablet,} \\
     & \multicolumn{1}{l}{Game Console, TV} \\
  \bottomrule
  \end{tabular}
\end{table}


\section{Characterizing Purchase Intent}
\label{section:characterizingpurchaseintent}

We explore customer behavior and, in particular, the difference in the behavior of purchasing and non-purchasing users. These explorations aim to identify characteristics that may help us improve the effectiveness of purchase intent predictors. We analyze several aspects of sessions, such as the length of purchase and non-purchase sessions, the temporal characteristics of sessions, and device information. Furthermore, we investigate the channels from which customers start sessions and issue queries during purchase sessions and non-purchase sessions. 

\subsection{Session Length}
\label{subsec:sess_len}

\begin{figure}[!htb]
\includegraphics[width=0.8\linewidth]{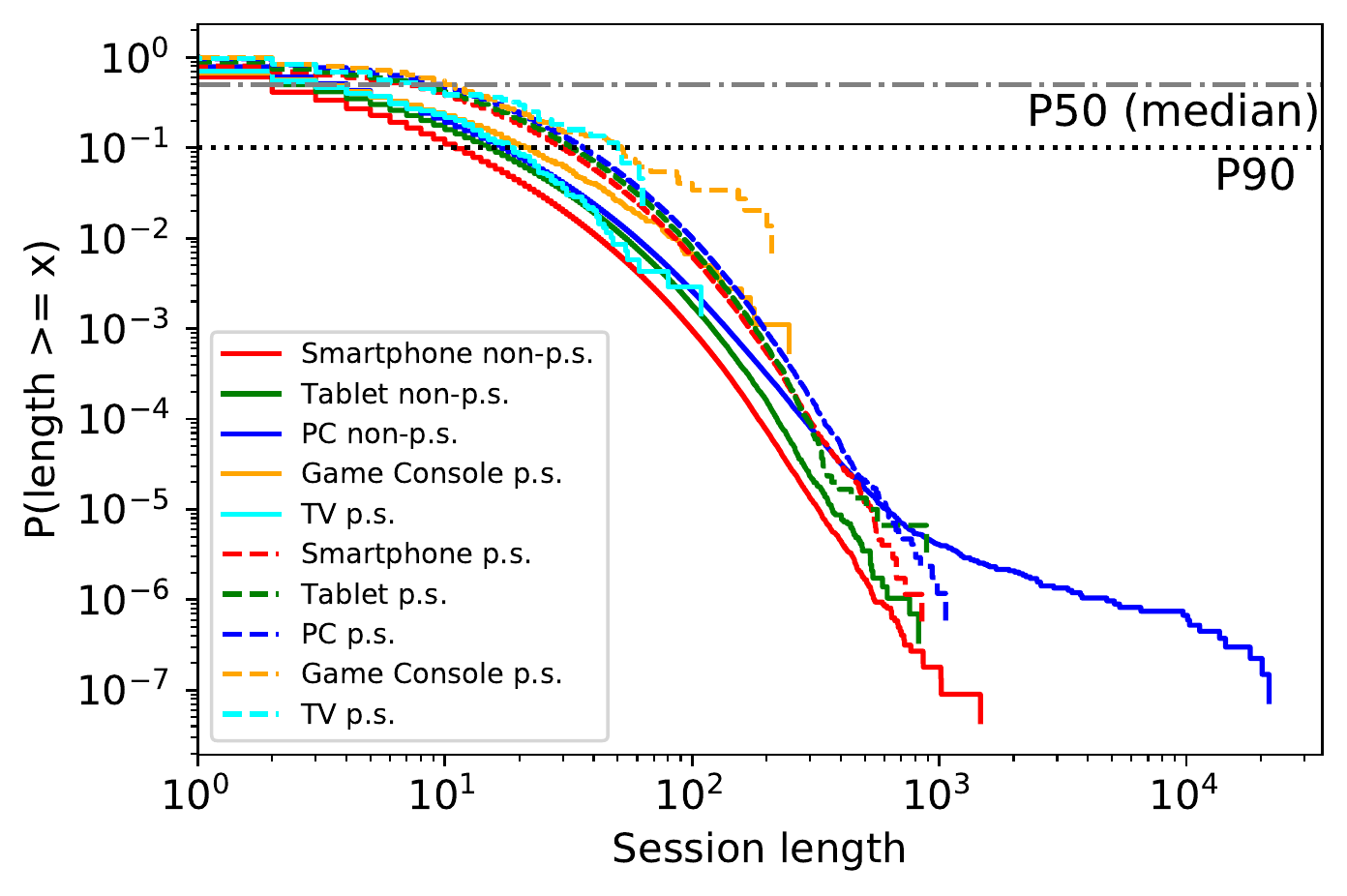}
\caption{CCDF of the session length per device type for purchase sessions (p.s.) and non-purchase sessions (non-p.s.).}
\label{fig:purchase_non_purchase_length}
\end{figure}

First, we examine the overall session length for purchase sessions and non-purchase sessions.
Figure~\ref{fig:purchase_non_purchase_length} plots the complementary cumulative distribution function (CCDF) of the session lengths of purchase sessions and non-purchase sessions per device type.

%

As can be seen in the area between the P50 and P90 percentiles in Figure~\ref{fig:purchase_non_purchase_length}, purchase sessions are in general longer than non-purchase sessions.
Moreover, the purchase session length per device varies less than the non-purchase session length per device. It can be explained by the fact that non-purchase sessions can be both very short or rather long, depending on the underlying user intents. For instance, a user could quickly look something up or spend some time exploring the catalog. On the other hand, in the case of purchase sessions, user intentions are less ambiguous.
Usually, users look for a specific product that they have in mind and, upon finding it, proceed to purchase.

From a device perspective, the shortest sessions take place on smartphones, whereas sessions on tablets are generally longer. The longest sessions occur on the PC, TV, and game console. This finding holds for both purchase sessions and non-purchase sessions.
However, in the tail of the distributions, the distinction between purchase session length and non-purchase session length is not as clear as between the P50 and P90 percentiles. The non-purchase session length distribution on the PC  has an exceptionally long tail.
Overall, we can attribute these findings to the fact that smartphones have a smaller screen and are therefore less convenient for longer sessions. Tablet screens are bigger than smartphone screens; hence, the sessions can last longer. The PC screen is the biggest one, and therefore PC users exhibit event longer sessions.


\subsection{Temporal Variations}

Next, we look into the temporal characteristics of purchase sessions and non-purchase sessions, such as their distribution across days of the week and the sessions' starting hours.

%
\begin{figure}[!htb]
\includegraphics[width=0.8\linewidth]{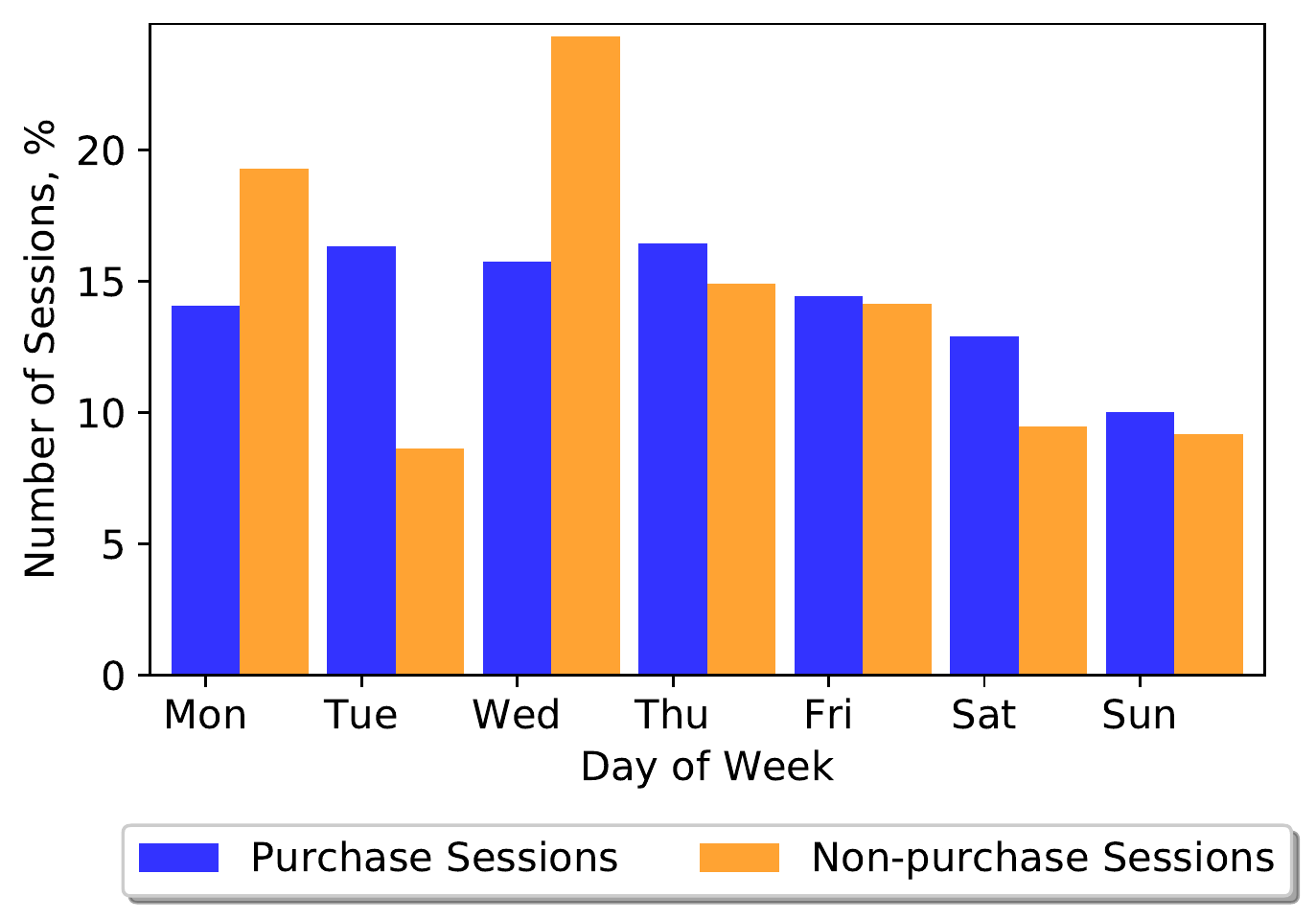}
\caption{The fraction of purchase sessions and non-purchase sessions across days of the week w.r.t. total amount of purchase and non-purchase sessions. Most activity occurs on weekdays.}
\label{fig:day_of_week_purchase_vs_non_purchase_sessions}
\end{figure}
%
First, we want to understand customer activity during the days of the week. Figure~\ref{fig:day_of_week_purchase_vs_non_purchase_sessions} shows how the number of purchase and non-purchase sessions varies across days of the week.
The three most popular days for purchase sessions are Thursday, Tuesday, and Wednesday. In total, the purchase sessions of these three days amount to $48.55\%$ of all purchase sessions.
On the other hand, the least popular purchase days are Sunday, Saturday, Monday, and Friday. They contribute to $51.45\%$ of purchase sessions.
The observed pattern of purchase behavior hints at the fact that customers prefer to buy during weekdays, which aligns with their workweek. Besides, we conclude that the lower purchase activity on Monday and Friday attributes to their proximity to weekends.

In the case of non-purchase sessions, the most active session days are Wednesday, Monday, and Thursday. Altogether, these days contribute to $58.55\%$ of non-purchase sessions. The least active days are Tuesday, Sunday, Saturday, and Friday. All the sessions of these days amount to $41.44\%$.
Just like for purchase sessions, the activity for non-purchase sessions also centers around weekdays.
However, the difference between the three most active days and the four least active days for non-purchase sessions is bigger than the corresponding difference for purchase sessions. For purchase sessions, the difference is only $2.9\%$, whereas, for non-purchase sessions, the difference is $17.11\%$. Moreover, Tuesday, the 
2-nd most popular day for purchase sessions is the least popular day for non-purchase sessions. On the other hand, Monday, the 2-nd most popular day for non-purchase sessions is the 3-rd least popular day for purchases.
The observation indicates that people need time to consider a purchase before making the buying decision. Hence, they spend Monday, the first day of the new week on considering the purchase, and the purchase itself happens on Tuesday or later in the week.
In general, the most active day of the week is Wednesday, whereas the least active day is Sunday.
These findings strongly suggest that user behavior depends on the day of the week. In general, people are most active on weekdays, during their workweek, their activity peaks in the middle of the week. On the other hand, at the beginning and end of the workweek, user activity is generally lower.

\begin{figure}[!h]
\includegraphics[width=0.8\linewidth]{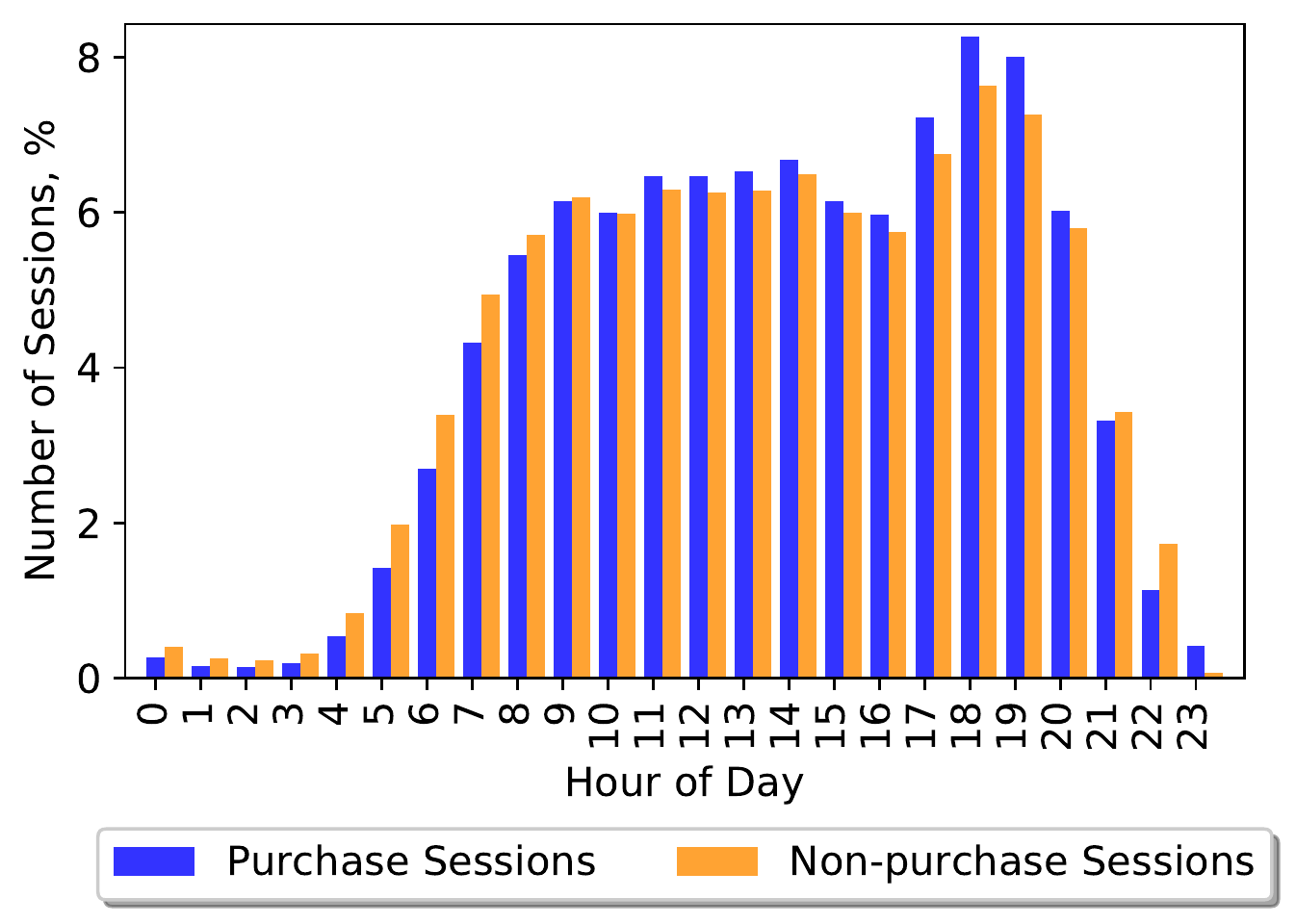}
\caption{The fraction of purchase session and non-purchase sessions across the hours of the day w.r.t. total amount of purchase and non-purchase sessions. Most purchase sessions start in the evening.}
\label{fig:hour_of_day_purchase_vs_non_purchase_sessions}
\end{figure}

Next, we look at user behavior on the level of the hour during which a session starts. As mentioned in Section~\ref{sec:dataset}, all the hours are represented in CET. Figure~\ref{fig:hour_of_day_purchase_vs_non_purchase_sessions} shows how purchase sessions and non-purchase sessions spread across the hours of the day.

%
%

As expected, the least active hours are in the early morning, in the period from 1 am to 3 am. That can be explained by the fact that most people sleep during the night. (Note that the majority of the e-commerce platform customers come from Europe; hence, time does not vary that much.)
Moreover, the activity on the platform during the period from 10 am till 5 pm is stable both for purchase and non-purchase sessions,
whereas the most active hours are in the evening, i.e., from 6 pm till 8 pm. In general, our observations correspond to the established rhythm of the daily life of the majority of people, who sleep during the night, browse e-commerce platforms both during work hours and in the evening after work.

\subsection{Channel Types}

Next, we look at whether channel types distributions change across purchase and non-purchase sessions. We define the following channel types: \textit{direct} where a user enters the platform directly; \textit{paid} where a user enters the platform through search engine advertisement, and \textit{organic} where a user enters the platform through a web search engine and unpaid results. Table~\ref{tab:channel_type} displays the channel distribution across purchase and non-purchase sessions.

\begin{table}[!h]
\caption{Channel types for purchase and non-purchase sessions.}
\label{tab:channel_type}
\begin{tabular}{l c c r}
\toprule
 & \multicolumn{2}{c}{\bf Sessions} & \bf Stand\\
\bf Channel & \bf Purchase (\%) & \bf Non-purchase (\%)  & \bf conv. rate \\ 
\midrule
Direct  &  \textbf{\numprint{71.07}}  & 
\textbf{\numprint{77.30}} & -0.56 \\
Paid & \numprint{16.74} 
&  \numprint{12.92}  & 0.54 \\
Organic & 
\numprint{11.78}  &  \phantom{0}\numprint{7.83} & \textbf{0.94} \\
Other & 
\phantom{0}\numprint{0.31} & 
\phantom{0}\numprint{1.05} & -1.33\\
\bottomrule
\end{tabular}
\end{table}
%

\noindent%
Both for purchase and non-purchase sessions, the direct channel is the most used channel to enter the platform. However, for purchase sessions, the percentage of sessions which start with the direct channel is $8.06\%$ less than the fraction of non-purchase sessions, which started with the direct channel.
The second most popular channel for purchase and non-purchase sessions is a paid channel. However, in the case of this channel, the fraction of purchase sessions is $12.92\%$ bigger than the corresponding channel type fraction for non-purchase sessions.
The organic channel is the third channel in terms of popularity for both session groups. The organic channel fraction for purchase sessions is $50.40\%$ bigger for purchase sessions when compared with non-purchase sessions.

Overall, during purchase sessions, users are more likely to enter the platform through paid or organic channels, whereas for non-purchase sessions the direct channel is more common. It can be explained by the fact that purchasers decide to converge after being offered an advertisement or a search result that matches their interest, whereas non-purchasers may enter the platform directly to explore the catalog.

\subsection{Devices}

In this subsection, we investigate purchase intent from the perspective of device types. In particular, we look at the device types used by purchasers and non-purchasers and analyze device switches.

\subsubsection{Device type}

First, we want to understand how many users are using multiple devices and which devices customers use for purchase and non-purchase sessions.

%
\begin{table}[!htb]
\caption{User device statistics per session.}
\label{tab:ds_user_device}
\begin{tabular}{lrr}
\toprule
\bf Device(s) & \bf Purchasers (\%) & \bf Non-purchasers (\%) \\
\midrule
> 1 device & 24.05 &  16.22      
\\
\midrule
1 device   &    75.95   
&   83.78 \\
2 devices   &  22.23 & 
15.39 \\
3 devices &   1.82 &  0.82 \\
4 devices & $\approx 0$ &  $\approx 0$ \\
5 devices & 0 & 0 \\
\bottomrule
\end{tabular}
\end{table}
%
\noindent%
Table~\ref{tab:ds_user_device} shows how many devices purchasers and non-pur\-chas\-ers own.
The majority of users from both groups are single-device users. However, the fraction of single-device purchasers is $9.35\%$ smaller than the corresponding fraction of non-purchasers.
On the other hand, the fraction of multi-device users for purchasers is $45.28\%$ bigger than the corresponding fraction for non-purchasers. In general, multi-device users represent almost a quarter of the purchasers.
As the number of devices increases, the difference between purchasers and non-purchasers grows.
Our observations support the statement that multi-device users tend to be more engaged~\citep{montanez2014cross}.

\begin{table}[!h]
\caption{Purchase and non-purchase sessions per device type and standardized conversion rates.}
\label{tab:ds_device_purchase}
\begin{tabular}{l @{} r r r}
\toprule
\multirow{2}{*}{\bf Device} & \bf Purchase & \bf Non-purchase & \bf Stand.\ \\
       &    \bf sessions (\%)     &    
\bf sessions (\%)  & \bf conv.\ rate \\
\midrule
Smartphone                        &    47.00\phantom{0}           & 58.09\phantom{0}     & $-$0.56 \\
PC                    &    44.97\phantom{0}              &   34.40\phantom{0} & 1.61     \\
Tablet                       
&   8.03\phantom{0}                &  7.50\phantom{0}    
&     0.61   \\
Game Console          
&    0.004     &   
0.004 & $-$0.40 \\
TV                              &    0.001   
&    0.002 & $-$1.25 \\
\bottomrule
\end{tabular}
\end{table}
%
Next, we examine the distributions of purchase and non-purchase sessions across device types and device-specific standardized conversion rates; see Table~\ref{tab:ds_device_purchase}.
%
The PC is the device with the highest conversion rate. Indeed, the fraction of purchase sessions is $30.70\%$ bigger than the fraction of non-purchase sessions.
The device with the second-highest conversion rate is a tablet. For this device, purchase sessions are $7.12\%$ more frequent than non-purchase sessions.
The Smartphone is the device with the second-lowest conversion rate. For this device, the number of purchase sessions is $19.10\%$ less frequent than the number of non-purchase sessions.

Game consoles and TVs are relatively new devices in e-commerce; hence, sessions with these devices are relatively less frequent.
Nevertheless, based on our observations, we find that the game console is a device with the third-highest conversion rate. Interestingly, its conversion rate is close to that of the smartphone. It can be explained by the fact that device functionalities of smartphones and tablets in e-commerce context blur due to the similarity of their interfaces and screen sized. The number of purchase sessions on a game console is $15.47\%$ less than the number of non-purchase sessions. The TV is the least common device, with the lowest conversion rate. The number of purchase sessions on this device is $34.01$ less than the number of non-purchase sessions.

We can explain our findings by the fact that customers use different devices for different purposes. For example, PCs and tablets seem to be used for the purchase, whereas smartphones, game consoles, and TVs for exploration.

\subsubsection{Device switches}

Next, we analyze how users switch between devices before a purchase session.

%
\begin{figure}[!h]
\includegraphics[width=0.9\linewidth]{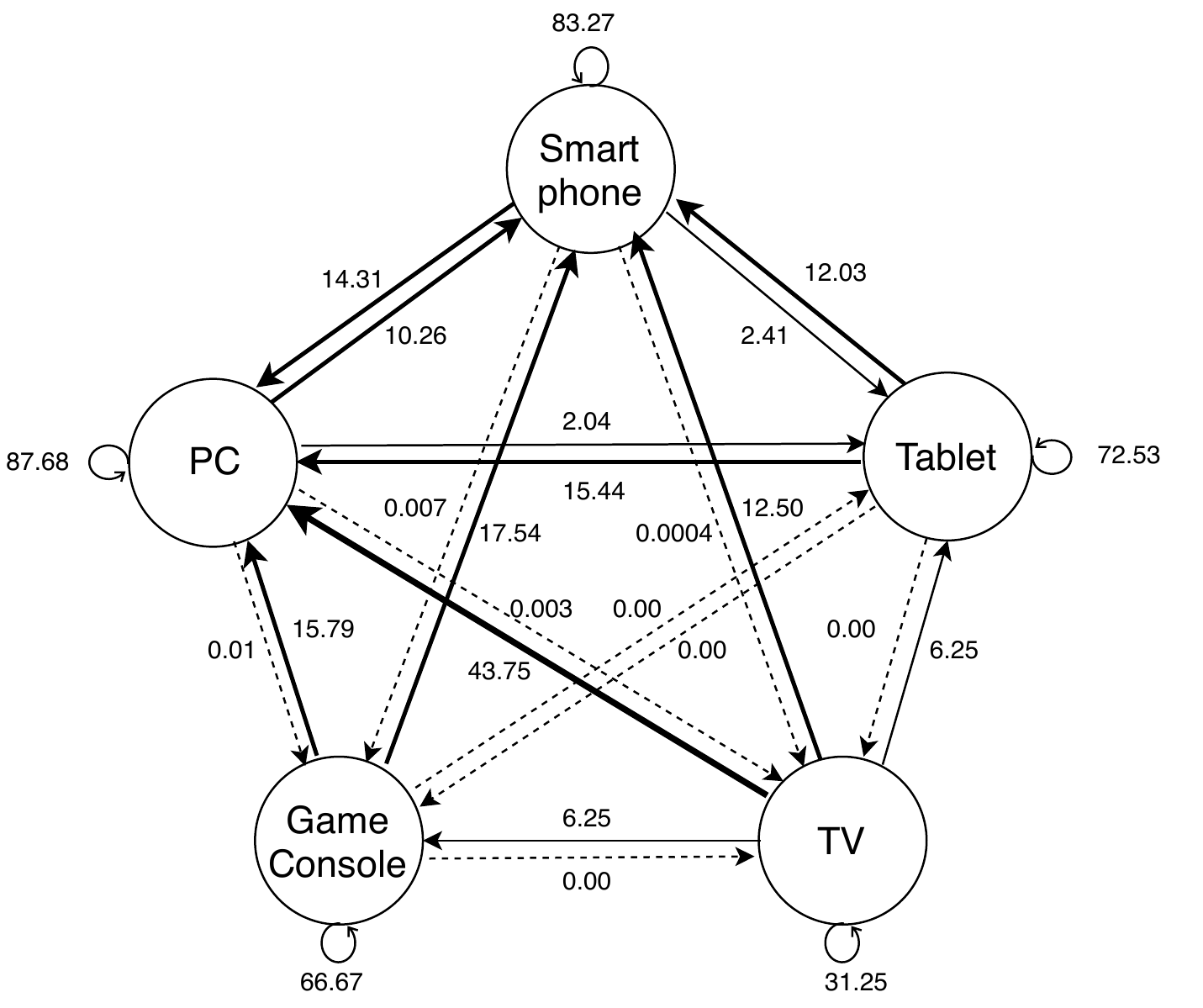}
\caption{Device transition probability before a purchase session, including self-transitions. The thickness of an arrow indicates the connection strength; the dashed line is the weakest connection.}
\label{fig:device_transitions_incl_self_transitions}
\end{figure}
Figure~\ref{fig:device_transitions_incl_self_transitions} shows device transition probability, including self-transi\-tions.
Generally, the situation when a user remains on the same device is the most likely outcome for all devices, except TV. There, the self-transition probability is lower than the probability of switching from TV to PC, a device with the highest self-transition probability. A probability of remaining on a smartphone is $5.03\%$ lower than a self-transition probability for PC, whereas a probability to remain on a tablet is $17.28\%$ lower than the probability to remain on PC. The game console has the second-lowest self-transition probability.

Next, we consider connections between two different devices. We characterize those interconnections based on how likely a user is to switch from one device to another one and vice versa.

\paragraph{Strong interconnections}
Some pairs of devices have high probability interconnections.
The strongest connection is between a smartphone and a PC, the two most popular devices.
The second strongest connection is between PC and tablet. There is a bigger discrepancy between probability rates, with the probability of switching to PC being $656.86\%$ higher than of switching to tablet.
The third strongest interconnection is between a smartphone and a tablet with a stronger connection switch to a smartphone, a more popular device. The probability of switching to a smartphone is $399.17\%$ higher than switching to a tablet.
Overall, the three interconnections form a triangle that includes the three most popular devices: PC, smartphone, and tablet.

\paragraph{One-sided interconnections}
For a one-sided interconnection there is a high probability of switching from one device to another, but a  close to zero probability of switching back. There are six cases of this type in Figure~\ref{fig:device_transitions_incl_self_transitions}.
TV is the device with the largest number of one-sided interconnections, with PC, smartphone, tablet, and game console. In all cases, the transition probability is low when TV is a target device, which can be explained by relative difficulty to purchase on TV.
The strongest one-sided interconnection is between TV and PC. The probability of switching from TV to PC is $43.75\%$, the highest transition probability for TV, and the highest probability to transition to another device. We explain this by the fact that PC is one of the most popular devices for purchase.
The second most likely device people switch to from TV is a smartphone, whereas a probability to switch to a tablet or game console is $6.25\%$.

Another device with a significant number of one-sided interconnections is the game console. Apart from the connection with TV discussed above, the device also has this connection type with smartphone and PC. The transition probability is close to zero when a game console is a target device. Unlike the situation with TV, came console has a higher probability of switching to a smartphone, whereas the probability of switching to PC is $1.75\%$ less.

In general, all one-sided interconnection cases include switching from a less common device type such as game console or TV to a more conventional device, such as PC, smartphone, or tablet.

\paragraph{Weak interconnections}
In some cases, the switch between two devices rarely happens, i.e., the transition probability is close to zero. As can be seen in Figure~\ref{fig:device_transitions_incl_self_transitions}, there is only one case of this type. It is a connection between a game console and a tablet.

\medskip\noindent%
Overall, the analysis of device switches before a purchase session supports the conclusion that users tend to switch from less popular devices such as TV and game console to more popular ones such as PC, smartphone, and tablet.

\subsection{Queries}

The next aspect of purchase intent that we examine is queries. We look at the number of queries in purchase and non-purchase sessions and per device type. In total, the dataset contains \numprint{31185176} queries, \numprint{1302195} or 4.17\% of which are unique. Given the number of sessions in the dataset, we can conclude that queries are infrequent.

%
\begin{table}[!h]
\caption{Queries per session for purchase and non-purchase sessions per device type, percentages are computed w.r.t. total number of queries per purchase or non-purchase session.}
\label{tab:ds_device_query}
\begin{tabular}{l @{} rr rr}
\toprule
\multirow{2}{*}{\bf Device} & \multicolumn{2}{c}{\bf Purchase sessions} & \multicolumn{2}{c}{\bf Non-purchase sessions}\\
\cmidrule(r){2-3}\cmidrule{4-5}
       &      \bf query/session     &    
\bf \%     &      \bf query/session      &    
\bf \%                                  
\\
\midrule
Smartphone  &  $ 4 $             &    \textbf{52.92}           &   $ 0.05$                      & 42.40 \\
PC                     & $ 2 $            &    36.38              &  $0.09$                            & \textbf{57.54} \\
Tablet                          &  $ 4 $    &   10.67               &  $0.0003$                                 &  0.049 \\
Game Console          &  $\approx 0$         &  $\approx 0$    &  $\approx 0$      & $\approx 0$  \\
TV                              &  $\approx 0$         &   $\approx 0$    &  $\approx 0$     &  $\approx 0$  \\
\midrule
Avg                              &  $3.16$       &    100              &    $0.06$            &   100 \\
\bottomrule
\end{tabular}
\end{table}
%

Table~\ref{tab:ds_device_query} shows the query per session frequencies across five devices for both purchase and non-purchase sessions. Besides, it also demonstrates which devices are most popular for querying during purchase and non-purchase sessions. Overall, queries are more common in purchase sessions. This can be explained by the fact that querying is more likely to happen when customers are determined to buy something.

Naturally, queries are most common for smartphones, PCs and tablets, and uncommon for game consoles and TVs. Indeed, the current interface of game console and TV makes it difficult to type queries, especially when compared to a PC or a smartphone.

The PC has the highest query per session frequency for non-purchase sessions and second-highest frequency for purchase sessions. A smartphone has the second-highest query per session frequency for non-purchase sessions and the highest query frequency per non-purchase session. Tablet, on the contrary, has the third-highest frequency for non-purchase sessions and the highest frequency for purchase sessions.

When it comes to query distributions per device for purchase and non-purchase sessions, the ranking is somewhat consistent for both groups. During purchase sessions, most queries are issued on a smartphone, whereas during non-purchase sessions PC prevails. On the other hand, PC is the second most popular device for purchase sessions, whereas for non-purchase sessions smartphone takes the second place. Tablet is third for both groups. In general, the query distribution across devices correlates with the session distribution across devices (see Table~\ref{tab:ds_device_purchase}).

%
\begin{table}[!h]
\caption{Unique query counts for purchase and non-purchase sessions per device type. The percentage is computed w.r.t. total number of queries per purchase or non-purchase session.}
\label{tab:ds_device_unique_query}
\begin{tabular}{l rr rr}
\toprule
\multirow{2}{*}{\bf Device} & \multicolumn{2}{c}{\bf Purchase sessions} & \multicolumn{2}{c}{\bf Non-purchase sessions}\\
\cmidrule(r){2-3}\cmidrule{4-5}
       &      \bf count     &    
\bf\%     &      \bf count     &    
\bf \%                                   \\
\midrule
Smartphone 
 &  \numprint{180542}                 &    36.89            &   \numprint{321640}                     & 39.57 \\
PC                     
& \numprint{224763}            
&    45.92              &  \numprint{312855}                             & 38.49 \\
Tablet                          &  \numprint{83881}            &   17.14                &  \numprint{175909}                                 &  21.64  \\
Game Console          
&  \numprint{136}         & 
0.03    &  \numprint{1841}      & 0.23 
\\
TV                              &  \numprint{46}          &   0.01   
&   \numprint{582}     & 
0.07  \\
\midrule
Total                              &   \numprint{489368}       &   
1.91              &    \numprint{812827}                             &   14.52 \\
\bottomrule
\end{tabular}
\end{table}

\noindent%
Next, we look at the number of unique queries for purchase and non-purchase sessions and per device. Table~\ref{tab:ds_device_unique_query} shows unique queries count and their corresponding fractions. The fractions are computed w.r.t. the total number of queries per session type and device. Overall, during purchase sessions users issue less unique queries, it holds for every device class but a PC. This can be explained by the fact that during purchase sessions users may retype a previous query to revisit the results they have seen earlier, whereas non-purchasers want to explore and hence use more unique queries.

\subsection{Purchase Intent Characteristics}
What have we learned from the log analysis conducted in this section that might help us to devise better models for purchase intent prediction?
We found out that purchase sessions tend to be longer what suggests that session length is an essential indicator of purchase intent. Besides, the difference in session length depends on the type of device customer use.
Moreover, we discovered how the day of week and hour of the day influence purchase behavior. In particular, customers are more likely to buy during the weekdays and in the evening.
From the perspective of channels, there is a difference, too. In particular, for non-purchase sessions, the direct channel is more common, whereas purchase sessions are more likely to start with paid or organic channels.
From the device perspective, we found out that multi-device users are more common among purchasers.
Besides, we figured the probability of purchase for every device and characterized transitions between devices.
After looking into queries in the dataset, we discovered that during purchase sessions, users issue more queries per session. Besides, during purchase session, there are less unique queries.


\section{Predicting Purchase Intent}
\label{section:predictingpurchaseintent}

Next, we turn to predict purchase intent when a user is anonymous (``anonymous setting'') and when a user is logged-in or recognized (``identified setting''). The goal of our experiments is to evaluate how the features which we discovered during dataset exploration influence purchase predictor performance in both settings. To accomplish this, we derive a feature set for each setting, and evaluate the features by adding them to an existing production-ready model, based on a Random Forest. To showcase the generalizability of our findings, we additionally test the impact of our features on five additional popular classifiers.
To investigate how the models' ability to predict purchase evolves throughout a session, we evaluate all models on 11 session steps (corresponding to the visits of 10 pages). We are interested in longer sessions because the outcome of such sessions is more difficult to predict. As we do not want to evaluate the model's performance on the very last step, (where the outcome is clear), we set up a buffer of 2 pages. Therefore, we filter out all the sessions which are shorter than 12 pages.
We conclude the section by analyzing the features which contributed most to the model performance in both the anonymous and the identified setting, and explore how dynamic and static feature importance change as the session continues.

\subsection{Experimental Setup}
\label{subsec:exper_setup}
In this section, we discuss the feature sets which we use in the experiments for the anonymous and identified setting, the models on which we test the features, and the evaluation setup.

\header{Feature sets} 
We start by designing a set of features for purchase prediction in identified and anonymous user settings. Since our initial analysis demonstrated that about 56\% of all sessions are anonymous (see Table~\ref{tab:ds_stats}), it is worth to pay special attention to this category. 
Based on the findings obtained thus far and on an analysis of best-performing features available in the literature \citep{hop2013web, lee2015online, niu2017predictive, seippel-2018-customer}, we compile a feature set presented in the Table~\ref{tab:feature_set}.

\begin{table}[!h]
\caption{Complete feature set. ``Dynamic'' indicates that a feature may change during a session.}
\label{tab:feature_set}
\begin{tabular}{llcc}
\toprule
\multicolumn{2}{l}{\bf Feature} & \bf Dynamic  & \bf Baseline  \\ 
\midrule
\multirow{8}{*}[-0.1cm]{\rotatebox{90}{Session}}
 & current page dwell time, mean & \checkmark & \checkmark\\
 & current page dwell time, $\sigma$ & \checkmark & \checkmark \\
 & page sequence score & \checkmark & \checkmark \\
 & number of pages & \checkmark & \checkmark\\
 & channel type \\
 & start hour \\
 & week day \\
 & device type \\
 & device conversion rate \\
\midrule
\multirow{9}{*}[0.5cm]{\rotatebox{90}{History}}
 & number of orders  &  & \checkmark \\
 & days since last purchase &  & \checkmark \\
 & number of sessions \\
 & number of devices \\
 & device sequence score \\
 & switch probability \\
\bottomrule
\end{tabular}
\end{table}

We categorize features into two classes: \emph{session features} and \emph{customer history features}. We derive session features from the information of the given session and base customer history features on the information from previous sessions of the given customer.

Since we run experiments in the anonymous and the identified setting, we use different feature sets for each setting. In the anonymous setting, the information about the customer is not available and, therefore, we can only use session features. On the other hand, when a customer is identified, we can use both session and customer history features.
The feature set contains both static and dynamic features. Dynamic features can change throughout the session, whereas static features remain constant.

\header{Models} Next, we select models on which we evaluate the features discovered during the dataset analysis. As our primary model, we use a production-ready classifier. This is a random forest (RF) with a baseline feature set as described in Table~\ref{tab:feature_set}.
Additionally, to showcase the general utility of our feature set, we experiment on additional models. After reviewing previous work in the domain of purchase prediction (see Section~\ref{sec:relatedwork}), we choose the following models for our experiments: logistic regression (LR), K-nearest neighbors (KNN), support vector machines (SVM), neural classifier, and gradient boosted decision tree (GBDT). Each model is trained on the baseline and extended feature set in both settings.

\header{Prediction setup} Since we want to explore how models' performances change across sessions, we select points of a session for which we predict the probability of purchase.

We define a point by the number of pages opened in the session up until the point of prediction. Overall, we select 11 points of measurement.  The first point is at the very beginning of the session when the user did not open any pages yet. At this point, the classifier makes a prediction based solely on static features. The following point of measurement is right after the user opened the first page. The subsequent nine points happen after the next nine pages. To make the evaluation possible and to ensure that we do not predict for the very last session page, we filter out sessions with fewer than 12 pages, with 2 pages as a buffer. The buffer is there to avoid the situation when the model predicts at the very end of a session when the outcome is clear. Therefore, we filter out all the sessions which are less than 12 pages long. For example, in step 2 we only have a session with at least 12 actions, which is a hard setting.

\header{Evaluation setup} For both settings, we evaluate model performance with 10-fold cross-validation. To account for class imbalance, we set class weights to be inversely proportional to class frequencies and use $F_{1}$ score as a primary evaluation metric.

\subsection{Prediction for Anonymous Users}

First, we evaluate how the added features influence model performance in the anonymous setting, where the user is not known.

\header{Setup} For the anonymous setting, we sampled \numprint{22982} sessions.  We use the data to create a feature set for the baseline model and our model. In the anonymous setting, there is no available information about customer history, therefore, we only use session features (see Table~\ref{tab:feature_set}. 
As can be seen from the table, the baseline feature set comprises four dynamic features, whereas the extended feature set offers five extra features. 
Since we predict purchase for different points in the session, we compute all dynamic features for a particular session point on which we evaluate. For each session point, we train the baseline and extended model on the obtained feature sets. 

\header{Results} The results in Figure~\ref{fig:results} show that the additional features boost model performance across all session steps. The performance boost is especially significant at step 0 when a customer has not opened any pages yet. In general, tree-based models (RF and GBDT) and the neural classifier demonstrate the highest scores across all steps. The models are followed by SVM, LR, and KNN classifiers.

The performance of all the models with the baseline feature set improved on step 1. The gain can be explained by the introduction of dynamic session features (step 0 means that the user did not open any pages yet, hence, no dynamic session features). Conversely, for models with the extended feature set, the introduction of dynamic features on step 1 does not significantly increase the performance. After step 1, models' performances reach a plateau.

\subsection{Prediction of Identified Users} 

Second, we test models' performances with baseline and extended feature sets in identified setting, when the user is known.

\header{Setup} For the identified setting, we sampled \numprint{6319} sessions.
The feature set for this setting includes session and customer history features (see Table~\ref{tab:feature_set}). 
During our experiments, we found out that information about the previous session device (such as device type and conversion rate) decrease model performance, so we excluded those features from the training and evaluation sets. This can be explained by the fact that the information about what kind of device users previously used and what was the probability of purchase on that device is not relevant for predicting purchase on the current device. 
In analogy with the anonymous setting, we prepare feature sets for each of the eleven session points and train and evaluate the models with the baseline and extended feature sets. 

\header{Results} Figure~\ref{fig:results} shows the performance of the models. Overall, the performance of all models for both baseline and extended feature sets and across all steps stays around 96\%. The only exception is the k-nearest neighbors classifier where adding extra features on step 0 increases the model's performance by 6.74\%. On step 1, however, the gain from the extended feature set is not present. This can be explained by the introduction of the dynamic session features.

\subsection{Feature Importance Analysis}

The experimental results raise a natural question that is 'Which features contribute most to model performance in both settings?' To answer this question, we look at the feature importance scores of a production-ready classifier which shows one of the best performances in both settings, random forest.

\begin{figure}
    \centering
    \includegraphics[height=4.2cm]{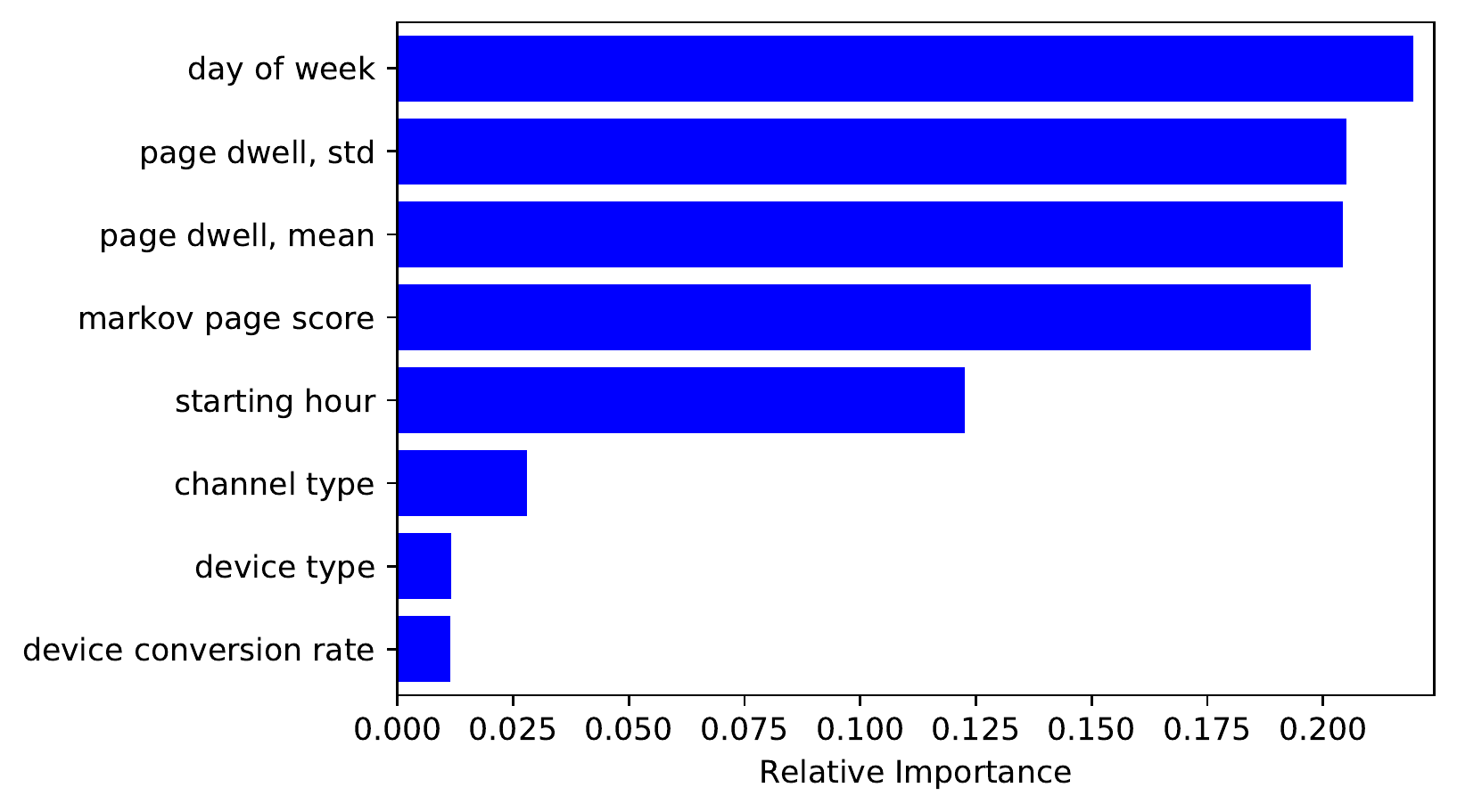}
    \includegraphics[height=4.2cm]{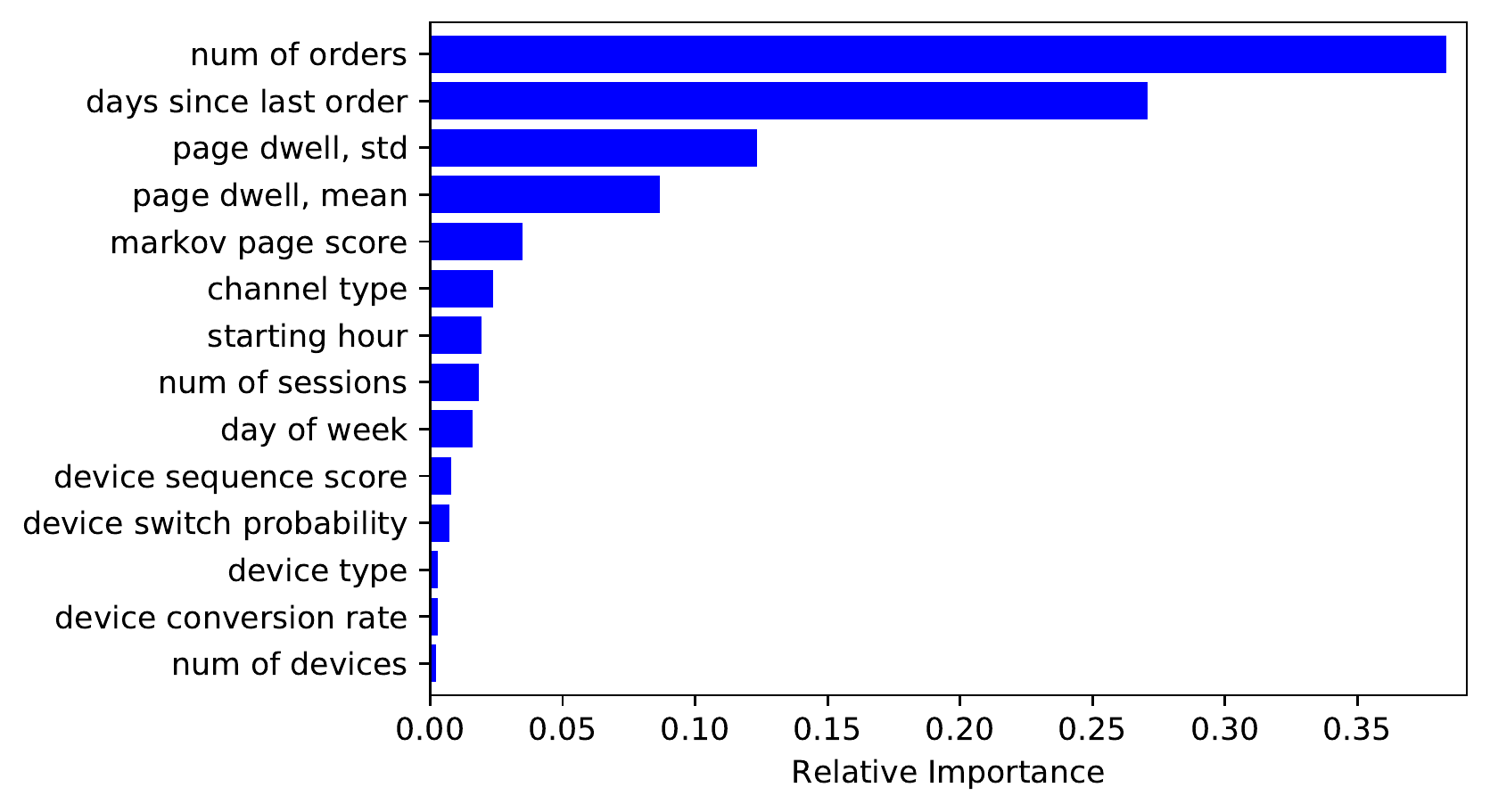}
    \mbox{}\hspace*{16.6mm}
    \includegraphics[clip,trim=0mm 0mm 0mm 10mm,height=4.18cm]{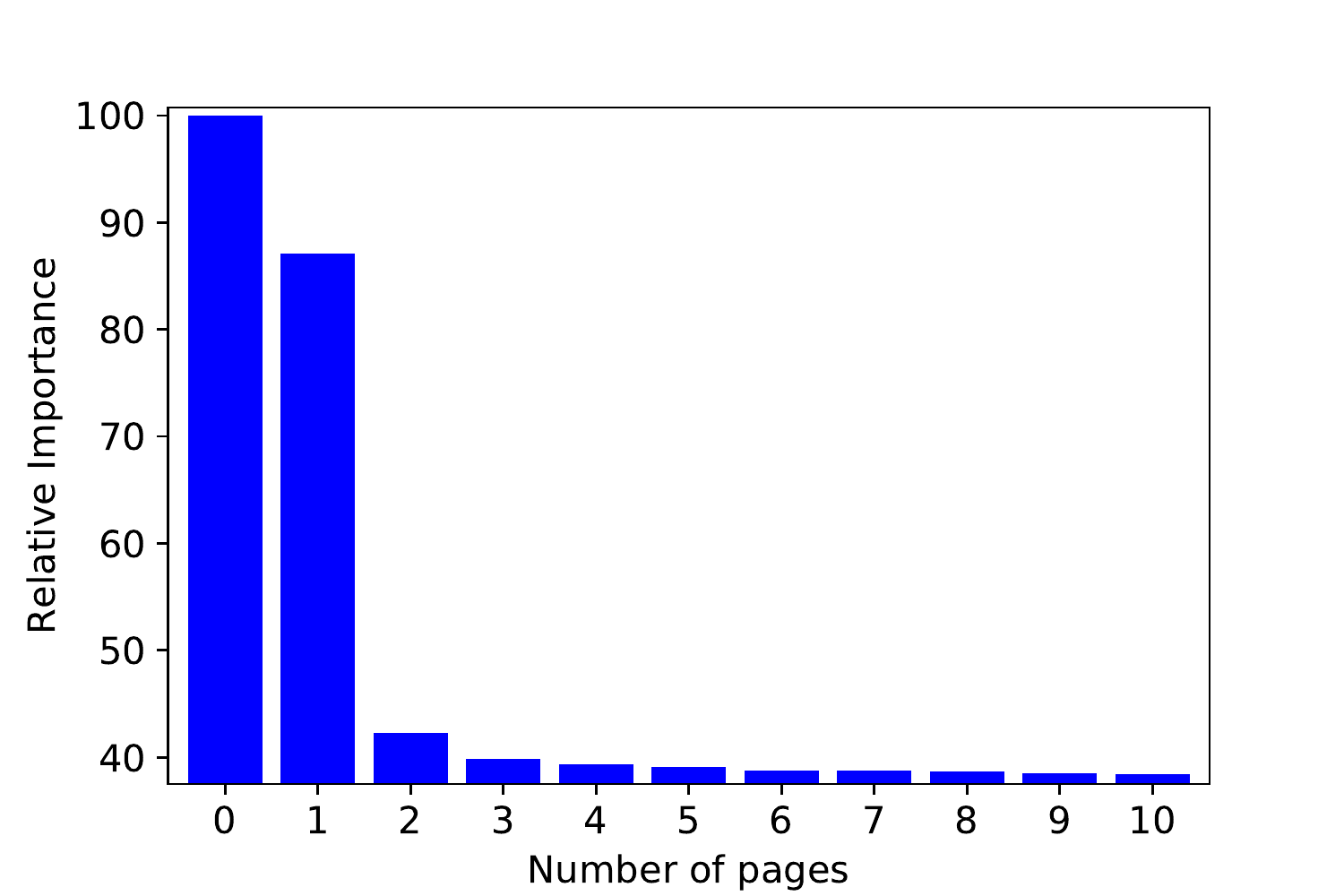}
    \caption{Feature importance for the Random Forest in the anonymous setting (top) and identified setting (center), as well as summed for static features in the anonymous setting (bottom).}
    \label{fig:evaluation}
    \label{fig:rf_anon_feat_imp}
    \label{fig:rf_iden_feat_imp}
    \label{fig:static_feature_importances}
\end{figure}

Figure~\ref{fig:rf_anon_feat_imp} (top) demonstrates that in anonymous setting, day of the week is the feature with the highest importance. It is followed by three dynamic features (standard deviation and mean of page dwell time, and Markov page sequence score), and four static features (starting hour, channel type, device type and conversion rate).

\begin{figure*}[!h]
\includegraphics[clip,trim=2mm 0mm 2mm 0mm,width=\linewidth]{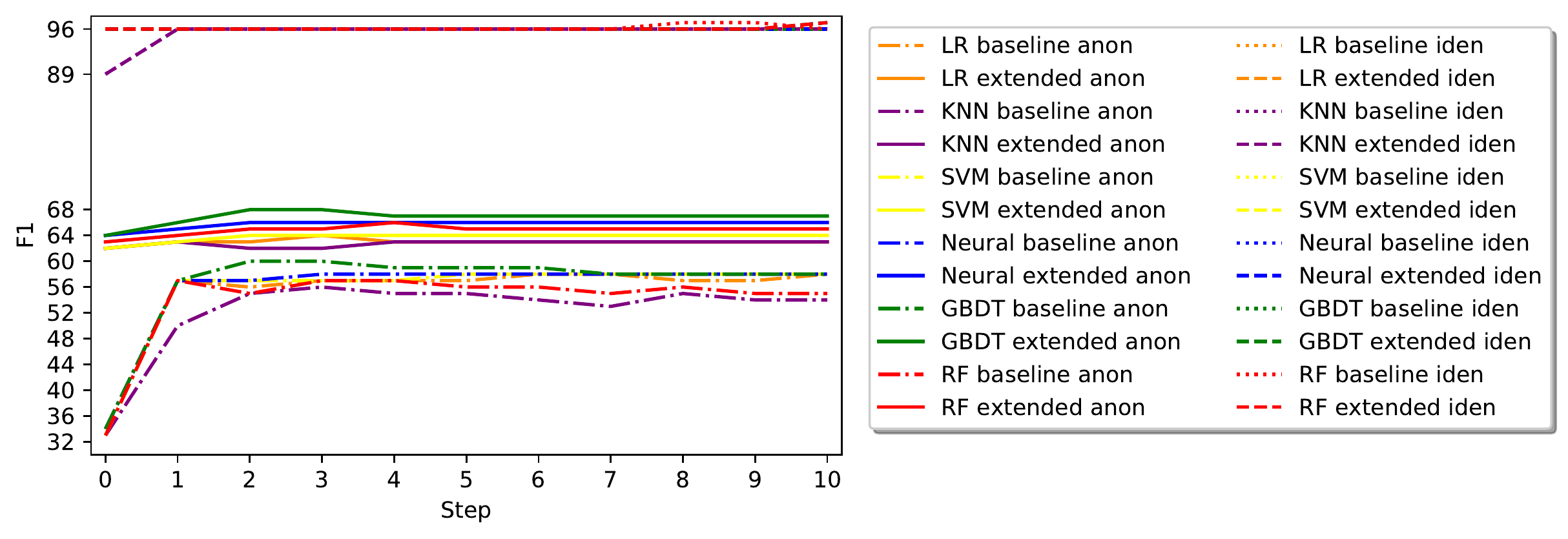}
\caption{Experimental results in anonymous (anon) and identified (iden) setting, across different session steps, $F_{1}$.}
\label{fig:results}
\end{figure*}

Figure~\ref{fig:rf_iden_feat_imp} (center) shows that in the identified user setting, number of previous orders, and number of days since last order are the features with the highest relative importance. Both features describe user historical purchasing behavior what can explain their high relative importance. 
The features are followed by three dynamic features (standard deviation and mean of page dwelling time, and Markov page sequence score), which also have relatively high importance in the anonymous setting.
The high relative importance of the dynamic session features (standard deviation and mean of page dwelling time, and Markov page sequence score) in both settings explain the gain all models with baseline feature set got on step 1 in the anonymous setting (see Figure~\ref{fig:results}).

Next, we determine how static feature importance changes across sessions. We consider the importance in the anonymous setting because the introduction of dynamic features in this setting showed an improvement.
Figure~\ref{fig:static_feature_importances} (bottom) shows that static session feature importance decrease as the session evolves, which entails that the importance of dynamic features increases. On step 0 the cumulative importance of static features is 100\% because there are no dynamic features introduced. However, from step 1 the relative importance starts to drop. The figure supports the hypothesis that as the session progresses dynamic features become more important.

\section{Related Work}
\label{sec:relatedwork}

\header{E-commerce user purchase behavior analysis} Research on understanding online users' purchasing behavior has been ongoing since the very beginning of e-commerce~\citep{bellman1999predictors}. Studies have investigated user motivation~\citep{bellman1999predictors}, factors that influence e-commerce adoption~\citep{o2003web}, as well as purchasing behavior~\citep{brown2003buying, hsu2006longitudinal}, with a focus on perceived security~\citep{salisbury2001perceived, george2002influences}, the decision-making process~\citep{senecal2005consumers}, and purchaser profiles~\citep{swinyard2004activities, hernandez2011age}. Besides, there has also been work on user behavior on content discovery platforms and its relationship to subsequent purchases~\citep{lo2016understanding} as well as work dedicated to the identification of a taxonomy of product search intents and the prediction of user satisfaction~\citep{su2018user}.

Unlike previous work, our study focuses on the exploration of user purchasing behavior by comparing purchase vs. non-purchase sessions. Besides, we analyze the data from the perspective of device types, and explore aspects such as session length, price of the seen products from the perspective of different devices. On top of that, we also look into the way customer switches between devices.

\header{Purchase prediction in e-commerce} The problem of e-commerce user behavior modeling has been studied from various angles, such as building multiple classifiers based on genetic algorithms~\cite{kim2003combination}, mining purchase patterns with association rules and using those patterns for purchase prediction~\cite{suh2004prediction}. Research has been focused on creating models robust to noise in session data~\cite{agichtein2006learning}, and using a recurrent neural network to predict customer behaviour~\cite{lang2017understanding}.

\citet{sismeiro2004modeling} predict purchasing task completion for a given user who completed at least one task earlier, whereas \citet{cheng2017predicting} explore user behavior on a content discovery platform to determine intent specificity and time in the future when a purchase is estimated to take place. Some work in the field focuses on using queries for purchasing behavior modeling. For instance, ~\citet{dai2006detecting} predict purchase based on input query. Besides using general session data, there has been work that incorporates demographic data and perceived attributes~\cite{young2004predicting}, scrolling and mouse movements~\cite{guo2010ready}, payment data~\cite{wen2018customer}, log-trace data~\cite{tao2019log2intent}, and phone touch actions~\cite{guo2019buying}. There has been work on analyzing behavioral patterns and the exploration of different model architectures. In particular, support vector machines, K-nearest neighbor approach, random forest, and logistic regression were used~\citep{lee2015online, suchacka2015k, niu2017predictive}.

Unlike previous work in this domain, our study focuses on purchase prediction with two types of users, identified and anonymous. Therefore, we develop two models, run them in two settings, and evaluate their results. The possibility to experiment with identified users also allows us to leverage information from previous user sessions, such as user purchasing history and the number of devices a user owns. In contrast, anonymous users contribute to a higher share of traffic, which makes it important to understand their behavior too. Additionally, we explore how the relevance of dynamic and static features changes as a session progresses.


\section{Discussion \& Conclusion}
\label{sec:discussionandimplications}

In this paper, we have carried out an analysis of user purchase intent in e-commerce. We have analyzed four weeks of session logs from a European e-commerce platform to identify signals in user behavior that can imply purchase intent. We have considered aspects such as session length, day of the week, and session start hour, as well as information about device, channel, and queries.


In the second part of our study, we have analyzed the relevance of the discovered signals by running a series of experiments aimed at purchase intent prediction in the anonymous and identified settings. We tested the features on random forest, the model which fits production requirements. Additionally, we tested the features on five other models. The experiments demonstrated the value of the features that we engineered based on our insights into the data. We explored which features contribute to performance improvement.

One of the implications of our study is enhanced understanding of purchasing user behavior in e-commerce. Understanding the behavior is the first step towards modeling it, as we demonstrated in the second part of the paper. 
Modeling user behavior can contribute towards reducing friction in the customer journey and, therefore, to better customer experience.
Besides, we explored the topic of detecting the purchase intent of anonymous users. We showed that, while anonymous users contribute to more than half of the traffic, their user intent is harder to detect because all the predictions have to be made without knowledge about the prior behavior.

Our research has several limitations; one of them is limited generalizability. Even though the data we use in our study comes from a dominant e-commerce platform, it is still only one platform. Hence, it would be interesting to verify the findings against other e-commerce platforms and explore the differences.
Moreover, we sampled four weeks of data, thereby introducing a sample bias that could make our findings sensitive to unknown temporal or seasonal patterns. Therefore, it would be interesting to explore if expanding our dataset will lead to new insights. For example, if we had several months of data, we could explore how user purchase intent changes across different months or seasons.
Furthermore, we evaluated our purchase intent prediction models in an offline setting. The next logical step is to evaluate them in an online setting.

Future research on the topic includes several directions.
First, there is an opportunity to continue research into general purchase behavior analysis and modeling in e-commerce. It would be interesting to explore more aspects of purchasing behavior and try out more models.
Another direction for further research concerns predicting purchase intent for anonymous users.
Another exciting direction for further research includes modeling device-specific purchase behavior. It can include both relatively common devices such as PC, smartphone, and tablet, and relatively less popular and studied devices such as TV or game console.

\section{Reproducibility}
All plots for our paper, as well as the code to regenerate them, can be found in our Git repository:\\ \url{https://github.com/mariyahendriksen/purchase_intent}.

\begin{acks}
This research was supported by Ahold Delhaize. We thank the three anonymous reviewers whose comments helped to improve and clarify the manuscript.
\end{acks}


\begin{thebibliography}{36}


\ifx \showCODEN    \undefined \def \showCODEN     #1{\unskip}     \fi
\ifx \showDOI      \undefined \def \showDOI       #1{#1}\fi
\ifx \showISBNx    \undefined \def \showISBNx     #1{\unskip}     \fi
\ifx \showISBNxiii \undefined \def \showISBNxiii  #1{\unskip}     \fi
\ifx \showISSN     \undefined \def \showISSN      #1{\unskip}     \fi
\ifx \showLCCN     \undefined \def \showLCCN      #1{\unskip}     \fi
\ifx \shownote     \undefined \def \shownote      #1{#1}          \fi
\ifx \showarticletitle \undefined \def \showarticletitle #1{#1}   \fi
\ifx \showURL      \undefined \def \showURL       {\relax}        \fi
\providecommand\bibfield[2]{#2}
\providecommand\bibinfo[2]{#2}
\providecommand\natexlab[1]{#1}
\providecommand\showeprint[2][]{arXiv:#2}

\bibitem[\protect\citeauthoryear{Agichtein, Brill, Dumais, and Ragno}{Agichtein
  et~al\mbox{.}}{2006}]%
        {agichtein2006learning}
\bibfield{author}{\bibinfo{person}{Eugene Agichtein}, \bibinfo{person}{Eric
  Brill}, \bibinfo{person}{Susan Dumais}, {and} \bibinfo{person}{Robert
  Ragno}.} \bibinfo{year}{2006}\natexlab{}.
\newblock \showarticletitle{Learning User Interaction Models for Predicting Web
  Search Result Preferences}. \bibinfo{booktitle}{\emph{SIGIR}}. 
\newblock


\bibitem[\protect\citeauthoryear{Bellman, Lohse, and Johnson}{Bellman
  et~al\mbox{.}}{1999}]%
        {bellman1999predictors}
\bibfield{author}{\bibinfo{person}{Steven Bellman}, \bibinfo{person}{Gerald
  Lohse}, {and} \bibinfo{person}{Eric~J Johnson}.}
  \bibinfo{year}{1999}\natexlab{}.
\newblock \showarticletitle{Predictors of Online Buying Behavior}.
\newblock \bibinfo{journal}{\emph{Commun. ACM}}  \bibinfo{volume}{42}
  (\bibinfo{year}{1999}), \bibinfo{pages}{32--48}.
\newblock


\bibitem[\protect\citeauthoryear{Ben-Shimon, Tsikinovsky, Friedmann, Shapira,
  Rokach, and Hoerle}{Ben-Shimon et~al\mbox{.}}{2015}]%
        {ben-shimon-2015-recsys}
\bibfield{author}{\bibinfo{person}{David Ben-Shimon},
  \bibinfo{person}{Alexander Tsikinovsky}, \bibinfo{person}{Michael Friedmann},
  \bibinfo{person}{Bracha Shapira}, \bibinfo{person}{Lior Rokach}, {and}
  \bibinfo{person}{Johannes Hoerle}.} \bibinfo{year}{2015}\natexlab{}.
\newblock \showarticletitle{RecSys Challenge 2015 and the YOOCHOOSE Dataset}.
  In \bibinfo{booktitle}{\emph{RecSys}}. \bibinfo{pages}{357--358}.
\newblock


\bibitem[\protect\citeauthoryear{Bomhardt, Gaul, and Schmidt-Thieme}{Bomhardt
  et~al\mbox{.}}{2005}]%
        {bomhardt-2005-web}
\bibfield{author}{\bibinfo{person}{Christian Bomhardt},
  \bibinfo{person}{Wolfgang Gaul}, {and} \bibinfo{person}{Lars
  Schmidt-Thieme}.} \bibinfo{year}{2005}\natexlab{}.
\newblock \showarticletitle{Web Robot Detection-preprocessing Web Logfiles for
  Robot Detection}.
\newblock In \bibinfo{booktitle}{\emph{CLADAG}}. \bibinfo{pages}{113--124}.
\newblock


\bibitem[\protect\citeauthoryear{Broder}{Broder}{2002}]%
        {broder-2002-taxonomy}
\bibfield{author}{\bibinfo{person}{Andrei Broder}.}
  \bibinfo{year}{2002}\natexlab{}.
\newblock \showarticletitle{A Taxonomy of Web Search}.
\newblock \bibinfo{journal}{\emph{SIGIR Forum}} \bibinfo{volume}{36},
  \bibinfo{number}{2}, \bibinfo{pages}{3--10}.
\newblock


\bibitem[\protect\citeauthoryear{Brown, Pope, and Voges}{Brown
  et~al\mbox{.}}{2003}]%
        {brown2003buying}
\bibfield{author}{\bibinfo{person}{Mark Brown}, \bibinfo{person}{Nigel Pope},
  {and} \bibinfo{person}{Kevin Voges}.} \bibinfo{year}{2003}\natexlab{}.
\newblock \showarticletitle{Buying or Browsing? An Exploration of Shopping
  Orientations and Online Purchase Intention}.
\newblock \bibinfo{journal}{\emph{EJM}}
  \bibinfo{volume}{37}, \bibinfo{number}{11/12} (\bibinfo{year}{2003}),
  \bibinfo{pages}{1666--1684}.
\newblock


\bibitem[\protect\citeauthoryear{Cheng, Lo, and Leskovec}{Cheng
  et~al\mbox{.}}{2017}]%
        {cheng2017predicting}
\bibfield{author}{\bibinfo{person}{Justin Cheng}, \bibinfo{person}{Caroline
  Lo}, {and} \bibinfo{person}{Jure Leskovec}.} \bibinfo{year}{2017}\natexlab{}.
\newblock \showarticletitle{Predicting Intent Using Activity Logs: How Goal
  Specificity and Temporal Range Affect User Behavior}. In
  \bibinfo{booktitle}{\emph{WWW}}.
\newblock


\bibitem[\protect\citeauthoryear{Dai, Zhao, Nie, Wen, Wang, and Li}{Dai
  et~al\mbox{.}}{2006}]%
        {dai2006detecting}
\bibfield{author}{\bibinfo{person}{Honghua~Kathy Dai}, \bibinfo{person}{Lingzhi
  Zhao}, \bibinfo{person}{Zaiqing Nie}, \bibinfo{person}{Ji-Rong Wen},
  \bibinfo{person}{Lee Wang}, {and} \bibinfo{person}{Ying Li}.}
  \bibinfo{year}{2006}\natexlab{}.
\newblock \showarticletitle{Detecting Online Commercial Intention}. In
  \bibinfo{booktitle}{\emph{WWW}}. ACM, \bibinfo{pages}{829--837}.
\newblock


\bibitem[\protect\citeauthoryear{George}{George}{2002}]%
        {george2002influences}
\bibfield{author}{\bibinfo{person}{Joey~F George}.}
  \bibinfo{year}{2002}\natexlab{}.
\newblock \showarticletitle{Influences on the Intent to Make Internet
  Purchases}.
\newblock \bibinfo{journal}{\emph{Internet Research}} \bibinfo{volume}{12},
  \bibinfo{number}{2} (\bibinfo{year}{2002}), \bibinfo{pages}{165--180}.
\newblock


\bibitem[\protect\citeauthoryear{Guo, Hua, Jia, Zhao, Wang, and Cui}{Guo
  et~al\mbox{.}}{2019}]%
        {guo2019buying}
\bibfield{author}{\bibinfo{person}{Long Guo}, \bibinfo{person}{Lifeng Hua},
  \bibinfo{person}{Rongfei Jia}, \bibinfo{person}{Binqiang Zhao},
  \bibinfo{person}{Xiaobo Wang}, {and} \bibinfo{person}{Bin Cui}.}
  \bibinfo{year}{2019}\natexlab{}.
\newblock \showarticletitle{Buying or Browsing?: Predicting Real-time
  Purchasing Intent using Attention-based Deep Network with Multiple Behavior}.
  \bibinfo{booktitle}{\emph{KDD}}.  \bibinfo{pages}{1984--1992}.
\newblock


\bibitem[\protect\citeauthoryear{Guo and Agichtein}{Guo and Agichtein}{2010}]%
        {guo2010ready}
\bibfield{author}{\bibinfo{person}{Qi Guo} {and} \bibinfo{person}{Eugene
  Agichtein}.} \bibinfo{year}{2010}\natexlab{}.
\newblock \showarticletitle{Ready to Buy or Just Browsing?: Detecting Web
  Searcher Goals from Interaction Data}. In
  \bibinfo{booktitle}{\emph{SIGIR}}.  \bibinfo{pages}{130--137}.
\newblock


\bibitem[\protect\citeauthoryear{Hern{\'a}ndez, Jim{\'e}nez, and
  Jos{\'e}~Mart{\'\i}n}{Hern{\'a}ndez et~al\mbox{.}}{2011}]%
        {hernandez2011age}
\bibfield{author}{\bibinfo{person}{Blanca Hern{\'a}ndez},
  \bibinfo{person}{Julio Jim{\'e}nez}, {and} \bibinfo{person}{M
  Jos{\'e}~Mart{\'\i}n}.} \bibinfo{year}{2011}\natexlab{}.
\newblock \showarticletitle{Age, Gender and Income: Do They Really Moderate
  Online Shopping Behaviour?}
\newblock \bibinfo{journal}{\emph{Online information review}}
  \bibinfo{volume}{35}, \bibinfo{number}{1}
  \bibinfo{pages}{113--133}.
\newblock


\bibitem[\protect\citeauthoryear{Hop}{Hop}{2013}]%
        {hop2013web}
\bibfield{author}{\bibinfo{person}{Walter Hop}.}
  \bibinfo{year}{2013}\natexlab{}.
\newblock \emph{\bibinfo{title}{Web-shop Order Prediction Using Machine
  Learning}}.
\newblock \bibinfo{thesistype}{Master's\ thesis}. \bibinfo{school}{Erasmus
  University Rotterdam}.
\newblock


\bibitem[\protect\citeauthoryear{Hsu, Yen, Chiu, and Chang}{Hsu
  et~al\mbox{.}}{2006}]%
        {hsu2006longitudinal}
\bibfield{author}{\bibinfo{person}{Meng-Hsiang Hsu}, \bibinfo{person}{Chia-Hui
  Yen}, \bibinfo{person}{Chao-Min Chiu}, {and} \bibinfo{person}{Chun-Ming
  Chang}.} \bibinfo{year}{2006}\natexlab{}.
\newblock \showarticletitle{A Longitudinal Investigation of Continued Online
  Shopping Behavior: An Extension of the Theory of Planned Behavior}.
\newblock \bibinfo{journal}{\emph{International Journal of Human-Computer
  Studies}} \bibinfo{volume}{64}, \bibinfo{number}{9} (\bibinfo{year}{2006}),
  \bibinfo{pages}{889--904}.
\newblock


\bibitem[\protect\citeauthoryear{Kim, Kim, and Lee}{Kim et~al\mbox{.}}{2003}]%
        {kim2003combination}
\bibfield{author}{\bibinfo{person}{Eunju Kim}, \bibinfo{person}{Wooju Kim},
  {and} \bibinfo{person}{Yillbyung Lee}.} \bibinfo{year}{2003}\natexlab{}.
\newblock \showarticletitle{Combination of Multiple Classifiers for the
  Customer's Purchase Behavior Prediction}.
\newblock \bibinfo{journal}{\emph{Decision Support Systems}}
  \bibinfo{volume}{34}, \bibinfo{number}{2} (\bibinfo{year}{2003}),
  \bibinfo{pages}{167--175}.
\newblock


\bibitem[\protect\citeauthoryear{Lang and Rettenmeier}{Lang and
  Rettenmeier}{2017}]%
        {lang2017understanding}
\bibfield{author}{\bibinfo{person}{Tobias Lang} {and} \bibinfo{person}{Matthias
  Rettenmeier}.} \bibinfo{year}{2017}\natexlab{}.
\newblock \showarticletitle{Understanding Consumer Behavior with Recurrent
  Neural Networks}. In \bibinfo{booktitle}{\emph{MLRec}}.
\newblock


\bibitem[\protect\citeauthoryear{Lee, Ha, Han, Rha, and Kwon}{Lee
  et~al\mbox{.}}{2015}]%
        {lee2015online}
\bibfield{author}{\bibinfo{person}{Munyoung Lee}, \bibinfo{person}{Taehoon Ha},
  \bibinfo{person}{Jinyoung Han}, \bibinfo{person}{Jong-Youn Rha}, {and}
  \bibinfo{person}{Ted~Taekyoung Kwon}.} \bibinfo{year}{2015}\natexlab{}.
\newblock \showarticletitle{Online footsteps to purchase: Exploring consumer
  behaviors on online shopping sites}. In \bibinfo{booktitle}{\emph{Proc. WebSci}}. \bibinfo{pages}{1--10}.
\newblock


\bibitem[\protect\citeauthoryear{Lo, Frankowski, and Leskovec}{Lo
  et~al\mbox{.}}{2016}]%
        {lo2016understanding}
\bibfield{author}{\bibinfo{person}{Caroline Lo}, \bibinfo{person}{Dan
  Frankowski}, {and} \bibinfo{person}{Jure Leskovec}.}
  \bibinfo{year}{2016}\natexlab{}.
\newblock \showarticletitle{Understanding Behaviors that Lead to Purchasing: A
  Case Study of Pinterest}. \bibinfo{booktitle}{\emph{KDD}}. \bibinfo{pages}{531--540}.
\newblock


\bibitem[\protect\citeauthoryear{Moe and Fader}{Moe and Fader}{2004}]%
        {moe2004dynamic}
\bibfield{author}{\bibinfo{person}{Wendy~W Moe} {and} \bibinfo{person}{Peter~S
  Fader}.} \bibinfo{year}{2004}\natexlab{}.
\newblock \showarticletitle{Dynamic Conversion Behavior at E-commerce Sites}.
\newblock \bibinfo{journal}{\emph{Management Science}} \bibinfo{volume}{50},
  \bibinfo{number}{3} (\bibinfo{year}{2004}), \bibinfo{pages}{326--335}.
\newblock


\bibitem[\protect\citeauthoryear{Montanez, White, and Huang}{Montanez
  et~al\mbox{.}}{2014}]%
        {montanez2014cross}
\bibfield{author}{\bibinfo{person}{George Montanez}, \bibinfo{person}{Ryen
  White}, {and} \bibinfo{person}{Xiao Huang}.} \bibinfo{year}{2014}\natexlab{}.
\newblock \showarticletitle{Cross-device Search}.
  \bibinfo{booktitle}{\emph{CIKM}}. \bibinfo{pages}{1669--1678}.
\newblock


\bibitem[\protect\citeauthoryear{Morwitz and Schmittlein}{Morwitz and
  Schmittlein}{1992}]%
        {morwitz1992using}
\bibfield{author}{\bibinfo{person}{Vicki~G Morwitz} {and}
  \bibinfo{person}{David Schmittlein}.} \bibinfo{year}{1992}\natexlab{}.
\newblock \showarticletitle{Using Segmentation to Improve Sales Forecasts Based
  on Purchase Intent: Which ``Intenders'' Actually Buy?}
\newblock \bibinfo{journal}{\emph{Journal of marketing research}}
  \bibinfo{volume}{29}, \bibinfo{number}{4} (\bibinfo{year}{1992}),
  \bibinfo{pages}{391--405}.
\newblock


\bibitem[\protect\citeauthoryear{Niu, Li, and Yu}{Niu et~al\mbox{.}}{2017}]%
        {niu2017predictive}
\bibfield{author}{\bibinfo{person}{Xi Niu}, \bibinfo{person}{Chuqin Li}, {and}
  \bibinfo{person}{Xing Yu}.} \bibinfo{year}{2017}\natexlab{}.
\newblock \showarticletitle{Predictive Analytics of E-commerce Search Behavior
  for Conversion}.
\newblock  (\bibinfo{year}{2017}).
\newblock


\bibitem[\protect\citeauthoryear{O'cass and Fenech}{O'cass and Fenech}{2003}]%
        {o2003web}
\bibfield{author}{\bibinfo{person}{Aron O'cass} {and} \bibinfo{person}{Tino
  Fenech}.} \bibinfo{year}{2003}\natexlab{}.
\newblock \showarticletitle{Web Retailing Adoption: Exploring the Nature of
  Internet Users Web Retailing Behaviour}.
\newblock \bibinfo{journal}{\emph{Journal of Retailing and Consumer services}}
  \bibinfo{volume}{10}, \bibinfo{number}{2} (\bibinfo{year}{2003}),
  \bibinfo{pages}{81--94}.
\newblock


\bibitem[\protect\citeauthoryear{Rowley}{Rowley}{2000}]%
        {rowley2000product}
\bibfield{author}{\bibinfo{person}{Jennifer Rowley}.}
  \bibinfo{year}{2000}\natexlab{}.
\newblock \showarticletitle{Product Search in E-shopping: A Review and Research
  Propositions}.
\newblock \bibinfo{journal}{\emph{Journal of Consumer Marketing}}
  \bibinfo{volume}{17}, \bibinfo{number}{1} (\bibinfo{year}{2000}),
  \bibinfo{pages}{20--35}.
\newblock


\bibitem[\protect\citeauthoryear{Salisbury, Pearson, Pearson, and
  Miller}{Salisbury et~al\mbox{.}}{2001}]%
        {salisbury2001perceived}
\bibfield{author}{\bibinfo{person}{W~David Salisbury},
  \bibinfo{person}{Rodney~A Pearson}, \bibinfo{person}{Allison~W Pearson},
  {and} \bibinfo{person}{David~W Miller}.} \bibinfo{year}{2001}\natexlab{}.
\newblock \showarticletitle{Perceived Security and World Wide Web Purchase
  Intention}.
\newblock \bibinfo{journal}{\emph{Industrial Management \& Data Systems}}
  \bibinfo{volume}{101}, \bibinfo{number}{4} (\bibinfo{year}{2001}),
  \bibinfo{pages}{165--177}.
\newblock


\bibitem[\protect\citeauthoryear{Seippel}{Seippel}{2018}]%
        {seippel-2018-customer}
\bibfield{author}{\bibinfo{person}{Hannah~Sophia Seippel}.}
  \bibinfo{year}{2018}\natexlab{}.
\newblock \emph{\bibinfo{title}{Customer Purchase Prediction through Machine
  Learning}}.
\newblock \bibinfo{thesistype}{Master's\ thesis}. \bibinfo{school}{University
  of Twente}.
\newblock


\bibitem[\protect\citeauthoryear{Senecal, Kalczynski, and Nantel}{Senecal
  et~al\mbox{.}}{2005}]%
        {senecal2005consumers}
\bibfield{author}{\bibinfo{person}{Sylvain Senecal}, \bibinfo{person}{Pawel~J
  Kalczynski}, {and} \bibinfo{person}{Jacques Nantel}.}
  \bibinfo{year}{2005}\natexlab{}.
\newblock \showarticletitle{Consumers' Decision-making Process and their Online
  Shopping Behavior: A Clickstream Analysis}.
\newblock \bibinfo{journal}{\emph{Journal of Business Research}}
  \bibinfo{volume}{58}, \bibinfo{number}{11} (\bibinfo{year}{2005}),
  \bibinfo{pages}{1599--1608}.
\newblock


\bibitem[\protect\citeauthoryear{Sismeiro and Bucklin}{Sismeiro and
  Bucklin}{2004}]%
        {sismeiro2004modeling}
\bibfield{author}{\bibinfo{person}{Catarina Sismeiro} {and}
  \bibinfo{person}{Randolph~E Bucklin}.} \bibinfo{year}{2004}\natexlab{}.
\newblock \showarticletitle{Modeling Purchase Behavior at an E-commerce Web
  Site: A Task-completion Approach}.
\newblock \bibinfo{journal}{\emph{Journal of marketing research}}
  \bibinfo{volume}{41}, \bibinfo{number}{3} (\bibinfo{year}{2004}),
  \bibinfo{pages}{306--323}.
\newblock


\bibitem[\protect\citeauthoryear{Su, He, Liu, Zhang, and Ma}{Su
  et~al\mbox{.}}{2018}]%
        {su2018user}
\bibfield{author}{\bibinfo{person}{Ning Su}, \bibinfo{person}{Jiyin He},
  \bibinfo{person}{Yiqun Liu}, \bibinfo{person}{Min Zhang}, {and}
  \bibinfo{person}{Shaoping Ma}.} \bibinfo{year}{2018}\natexlab{}.
\newblock \showarticletitle{User Intent, Behaviour, and Perceived Satisfaction
  in Product Search}. \bibinfo{booktitle}{\emph{WSDM}}. 
  \bibinfo{pages}{547--555}.
\newblock


\bibitem[\protect\citeauthoryear{Suchacka, Skolimowska-Kulig, and
  Potempa}{Suchacka et~al\mbox{.}}{2015}]%
        {suchacka2015k}
\bibfield{author}{\bibinfo{person}{Gra{\.z}yna Suchacka},
  \bibinfo{person}{Magdalena Skolimowska-Kulig}, {and} \bibinfo{person}{Aneta
  Potempa}.} \bibinfo{year}{2015}\natexlab{}.
\newblock \showarticletitle{A k-Nearest Neighbors Method for Classifying User
  Sessions in E-commerce Scenario}.
\newblock \bibinfo{journal}{\emph{Journal of Telecommunications and Information
  Technology}} (\bibinfo{year}{2015}).
\newblock


\bibitem[\protect\citeauthoryear{Suh, Lim, Hwang, and Kim}{Suh
  et~al\mbox{.}}{2004}]%
        {suh2004prediction}
\bibfield{author}{\bibinfo{person}{Euiho Suh}, \bibinfo{person}{Seungjae Lim},
  \bibinfo{person}{Hyunseok Hwang}, {and} \bibinfo{person}{Suyeon Kim}.}
  \bibinfo{year}{2004}\natexlab{}.
\newblock \showarticletitle{A Prediction Model for the Purchase Probability of
  Anonymous Customers to Support Real Time Web Marketing}.
\newblock \bibinfo{journal}{\emph{Expert Systems with Applications}}
  \bibinfo{volume}{27}, \bibinfo{number}{2}
  \bibinfo{pages}{245--255}.
\newblock


\bibitem[\protect\citeauthoryear{Swinyard and Smith}{Swinyard and
  Smith}{2004}]%
        {swinyard2004activities}
\bibfield{author}{\bibinfo{person}{William~R Swinyard} {and}
  \bibinfo{person}{Scott~M Smith}.} \bibinfo{year}{2004}\natexlab{}.
\newblock \showarticletitle{Activities, Interests, and Opinions of Online
  Shoppers and Non-shoppers}.
\newblock \bibinfo{journal}{\emph{IBER}} \bibinfo{volume}{3}, \bibinfo{number}{4}
  (\bibinfo{year}{2004}).
\newblock


\bibitem[\protect\citeauthoryear{Tao, Li, Wang, Fang, Yang, Zhao, and Fu}{Tao
  et~al\mbox{.}}{2019}]%
        {tao2019log2intent}
\bibfield{author}{\bibinfo{person}{Zhiqiang Tao}, \bibinfo{person}{Sheng Li},
  \bibinfo{person}{Zhaowen Wang}, \bibinfo{person}{Chen Fang},
  \bibinfo{person}{Longqi Yang}, \bibinfo{person}{Handong Zhao}, {and}
  \bibinfo{person}{Yun Fu}.} \bibinfo{year}{2019}\natexlab{}.
\newblock \showarticletitle{Log2Intent: Towards Interpretable User Modeling via
  Recurrent Semantics Memory Unit}. \bibinfo{booktitle}{\emph{KDD}}. \bibinfo{pages}{1055--1063}.
\newblock


\bibitem[\protect\citeauthoryear{Tsagkias, King, Kallumadi, Murdock, and
  de~Rijke}{Tsagkias et~al\mbox{.}}{2020}]%
        {tsagkias-2020-challenges}
\bibfield{author}{\bibinfo{person}{Manos Tsagkias},
  \bibinfo{person}{Tracy~Holloway King}, \bibinfo{person}{Surya Kallumadi},
  \bibinfo{person}{Vanessa Murdock}, {and} \bibinfo{person}{Maarten de Rijke}.}
  \bibinfo{year}{2020}\natexlab{}.
\newblock \showarticletitle{Challenges and Research Opportunities in eCommerce
  Search and Recommendations}.
\newblock \bibinfo{journal}{\emph{SIGIR Forum}} \bibinfo{volume}{54},
  \bibinfo{number}{1} (\bibinfo{date}{June} \bibinfo{year}{2020}).
\newblock


\bibitem[\protect\citeauthoryear{Wen, Yeh, Tsai, Peng, and Shuai}{Wen
  et~al\mbox{.}}{2018}]%
        {wen2018customer}
\bibfield{author}{\bibinfo{person}{Yu-Ting Wen}, \bibinfo{person}{Pei-Wen Yeh},
  \bibinfo{person}{Tzu-Hao Tsai}, \bibinfo{person}{Wen-Chih Peng}, {and}
  \bibinfo{person}{Hong-Han Shuai}.} \bibinfo{year}{2018}\natexlab{}.
\newblock \showarticletitle{Customer Purchase Behavior Prediction from Payment
  Datasets}. In \bibinfo{booktitle}{\emph{WSDM}}. \bibinfo{pages}{628--636}.
\newblock


\bibitem[\protect\citeauthoryear{Young~Kim and Kim}{Young~Kim and Kim}{2004}]%
        {young2004predicting}
\bibfield{author}{\bibinfo{person}{Eun Young~Kim} {and}
  \bibinfo{person}{Youn-Kyung Kim}.} \bibinfo{year}{2004}\natexlab{}.
\newblock \showarticletitle{Predicting Online Purchase Intentions for Clothing
  Products}.
\newblock \bibinfo{journal}{\emph{EJM}}
  \bibinfo{volume}{38}, \bibinfo{number}{7} (\bibinfo{year}{2004}),
  \bibinfo{pages}{883--897}.
\newblock


\end{thebibliography}


\end{document}